\documentclass[12pt]{iopart}

\usepackage{graphicx} 
\usepackage{picture} 
\usepackage{epstopdf} 
\usepackage{pdfpages} 
\usepackage{microtype} 
\usepackage{tikz} 
\usepackage{xspace}
\usepackage[section]{placeins}
\usepackage[shortcuts]{extdash}
\usepackage{hyperref}
\usepackage{iopams}
\usepackage{amstext} 
\usepackage{stackrel} 

\begin{document}

\title[Tunable self-assembled spin chains of strongly interacting cold
atoms]{Tunable self-assembled spin chains of strongly interacting cold atoms
for demonstration of reliable quantum state transfer}
\date{\today}

\author{N~J~S Loft$^1$, O~V Marchukov$^1$, D Petrosyan$^{2}$, and
  N~T Zinner$^1$}

\address{$^1$ Department of Physics and Astronomy,
  Aarhus University, DK-8000 Aarhus C,  Denmark}
\address{$^2$ Institute of Electronic Structure and Laser, FORTH, 
GR-71110 Heraklion, Crete, Greece}

\begin{abstract}
  We have developed an efficient computational method to treat long, 
  one-dimensional systems of strongly-interacting atoms forming
  self-assembled spin chains. Such systems can be used to realize
  many spin chain model Hamiltonians tunable by the external 
  confining potential. As a concrete demonstration, we consider 
  quantum state transfer in a Heisenberg spin chain and we show 
  how to determine the confining potential in order to obtain
  nearly-perfect state transfer.
\end{abstract}



\submitto{\NJP}

\maketitle

\section{Introduction}

Recent experimental progress with cold atomic gases has opened the
door for realizing quantum many-body systems and simulating some of
the most useful models of many-body physics
\cite{lewenstein2007,bloch2008,esslinger2010,baranov2012,zinner2013}.
Over the last decade, it has become possible to perform experiments
with effective one-dimensional few- and many-body systems
\cite{moritz2003,stoferle2004,kinoshita2004,paredes2004,kinoshita2006,%
  haller2009,haller2010,serwane2011,zurn2012,wenz2013,pagano2014,murmann2015}.
Using Feshbach resonances \cite{chin2010}, one can achieve tunable
interactions in one-dimensional traps \cite{olshanii1998} and reach
the regime of strongly interacting particles.  Combining optical and
magnetic traps and standing-wave optical lattices, one may realize
different spin models with cold atoms \cite{kuklov2003,duan2003} 
and thus study the static properties and non-equilibrium dynamics 
of quantum magnetism \cite{fukuhara2013,hild2014}.  In particular, 
one can realize the paradigmatic Heisenberg spin chain using cold atoms 
with strong interactions \cite{hild2014}.

The local exchange coefficients entering the effective Heisenberg 
spin chain model are strongly dependent on the geometry 
of the atom trap. Computing the relation between the trap 
parameters and the exchange coefficients would provide the
essential link between theory and experiment for specific
implementations of numerous theoretical spin chain models with
strongly interacting cold atomic gases. A method that provides such 
a link was developed \cite{volosniev2014,deuretzbacher2014}, but 
the computation of the nearest-neighbor interaction coefficients 
for a system of more than a few atoms was a daunting task. 
Recently, these computational difficulties were overcome 
\cite{loft2016b,loft2015,Deuretzbacher2016} and we can now efficiently
treat long one-dimensional systems and obtain self-assembled spin
chain Hamiltonians tunable by the external trap potential. Engineering
the spin chains in this way allows us to study the delicate interplay
between the strong atom-atom interactions and the trap parameters,
gaining important theoretical insight into the physics of the experiment.

As a demonstration of our method \cite{loft2016b}, we will focus here on 
a specific yet important problem of excitation transfer in spin chains. 
Within the field of quantum information, a key subject is the study 
of coherent state and excitation transfer properties of quantum systems
\cite{StrRevs2007,nikolopoulos2014}. Typically, a quantum computation
system is assembled from a number of smaller two-level systems (qubits) 
interacting with each other and/or with the surrounding environment. 
Quantum information can be encoded locally by, for instance, 
preparing an appropriate superposition of spin-up and spin-down states 
of one of the qubits. As time evolves, this information will propagate
through the system and can potentially be retrieved somewhere else. 
The ability to transfer such quantum information reliably from one point 
to another is of great importance in quantum information theory. 
The pioneering work of Bose \cite{bose2003} showed that spin chains 
are promising candidates for realizing a ``quantum bus'' that can 
implement this quantum information transfer. 

In 2004 it was shown that one could obtain perfect state transfer in 
a one-dimensional $N$-particle spin chain by appropriately tuning the 
local exchange coefficients \cite{christandl2004, nikolopoulos2004,Plenio2004}. 
To the best of our knowledge, so far the perfect state transfer protocol
was experimentally demonstrated only in the system of evanescently 
coupled optical waveguides \cite{Bellec2012,Perez-Leija2013}. 
This is in fact a classical system that can emulate only single 
particle (non-interacting) quantum physics, since the coupled-mode 
equations for a waveguide array are equivalent to the amplitude 
equations for the single particle (excitation) dynamics in a 
lattice (chain). It is therefore important to propose and analyze 
a real many-body quantum system for realizing properly engineered 
spin chains. 

In this paper, we consider a cold-atom implementation of such a system.
We demonstrate our efficient computation method by investigating how the 
trap potential can be chosen in order to obtain perfect state transfer 
for up to $N=20$ particles. 
Furthermore, we consider how sensitive is the state transfer to 
the experimentally realistic noise. The noise in our calculations
is included exactly in the sense that we add fluctuations to the external 
potential and then compute the local exchange coefficients. This should 
be contrasted with first approximating the true continuum Hamiltonian 
by a lattice model in the strongly interacting limit, and then adding 
homogeneous noise to the coefficients of the lattice model. Here we 
abstain from the approximate lattice description and include the noise 
already in the continuum model from which we compute the corresponding 
spin chain parameters and their uncertainties. We find that high-fidelity 
state transfer can tolerate moderate amount of noise, thus suggesting 
that the cold-atom implementations of truly many-body quantum systems for 
quantum information applications are well-suited for experimental study.

The paper is organized as follows: In Section~\ref{sec:spinchain} we
briefly explain how the strongly interacting cold-atom gas is related
to the spin chain system and sketch the solution for this general
problem.  In Section~\ref{sec:perfect} we introduce the particular spin
chain system exhibiting perfect state transfer properties, and we find
the nearly optimal form of the confining potential for the cold-atom
gas. In Section~\ref{sec:noise} we study experimentally realistic 
fluctuations of the trapping potential and show that the transfer 
is tolerant to a moderate amount of the resulting noise. 
We close the paper by summarizing our results.

\section{The general spin chain problem}
\label{sec:spinchain}
We begin with an outline of the general spin chain problem and its
connection to strongly interacting one-dimensional gases, summarizing
some important results from Ref.~\cite{volosniev2015} (for related studies 
that also map the strongly-interacting systems onto spin models see 
Refs.~\cite{deuretzbacher2014,Deuretzbacher2016,pu2015,levinsen2015,hu2015,yang2015}).
The general problem concerns $N$ strongly-interacting particles with
mass $m$ in an arbitrary one-dimensional confining potential $V(x)$.
We consider two species (spin states) of particles, denoting by
$N_\uparrow$ the number of spin-up particles and by $N_\downarrow = N
- N_\uparrow$ the number of spin-down particles.  Particles of
different species interact via a contact potential with strength $g > 0$ 
and of the same species with strength $\kappa g$ with $\kappa > 0$. 
The Hamiltonian of the system is
\begin{eqnarray}
  \label{eq:hamiltonian}
  H = &\sum\limits_{i=1}^N
  \left[ - \frac{\hbar^2}{2m} \frac{\partial^2}{\partial x_i^2}
         + V(x_i) \right]
  + g \sum\limits_{\uparrow\downarrow}
  \delta(x_i - x_j) \nonumber\\
  &+\kappa g \sum\limits_{\uparrow\uparrow}
  \delta(x_i - x_j)
  +\kappa g \sum\limits_{\downarrow\downarrow}
  \delta(x_i - x_j)
  \; ,
\end{eqnarray}
where the sum over $\uparrow\downarrow$ is understood as a sum 
over the $N_\uparrow \cdot N_\downarrow$ pairs of particles of different
species, while the sum over $\uparrow\uparrow$ is over all 
$N_\uparrow\cdot(N_\uparrow-1)/2$ pairs of spin-up particles,
and similarly for $\downarrow\downarrow$. As noted in
Ref.~\cite{volosniev2015}, the Hamiltonian of
Eq.~\eref{eq:hamiltonian} is valid for both bosons and fermions.

For strong interaction $g \gg 1$ and $\kappa \gg 1/g$, we can solve
for the system described by Eq.~\eref{eq:hamiltonian} exactly, to first
order in $1/g$, if we know the $N$ lowest-energy eigenstates of the
single-particle Hamiltonian
\begin{equation}
  \label{eq:single-particle-hamiltonian}
  H_1(x) = -\frac{\hbar^2}{2m} \frac{\partial^2}{\partial x^2} + V(x) \; .
\end{equation}
If the trap potential $V(x)$ is sufficiently well-behaved, solving the
Schr\"odinger equation $H_1 \psi_i = E_i \psi_i$ numerically is
an elementary exercise. Since we are interested in strong
interactions, we perturb the system away from the Tonks-Girardeau (TG) 
limit $1/g \rightarrow 0$. In this limit, the ground state is
$N!/(N_\uparrow! \cdot N_\downarrow!)$-fold degenerate. We denote this
TG ground state energy by $E_0$, which is also the energy of the 
(non-normalized) Slater determinant wave function
$\Phi_0(x_1,\dots,x_N)$ composed of the $N$  
single-particle eigenfunctions $\psi_i(x)$.
From now on we will use the units of $\hbar=m=1$.

Moving slightly away from the TG limit, the ground-state degeneracy is
lifted, and the Hamiltonian can be mapped onto a Heisenberg spin chain
model. To linear order in $1/g \ll 1$, the Hamiltonian can be written as
\begin{equation}
  \label{eq:heisenberg}
  H = E_0 - \sum_{k=1}^{N-1} \frac{\alpha_k}{g} \left[ \frac{1}{2}
    \left(1 - \boldsymbol{\sigma}^k \cdot \boldsymbol{\sigma}^{k+1} \right)
    + \frac{1}{\kappa}
    \left(1 + \sigma_z^k \sigma_z^{k+1} \right) \right] \; ,    
\end{equation}
where $\boldsymbol{\sigma}^k = (\sigma_x^k, \sigma_y^k, \sigma_z^k)$
are the Pauli matrices acting on the spin of the particle at $k$th position, 
and $\alpha_k$ are the coefficients determined solely by the trapping potential 
$V(x)$ \cite{volosniev2015}. In other words, the geometric coefficients
$\alpha_k$ only depend on the \emph{geometry} of the trap. Calculation 
of all the geometric coefficients for a given trap potential would determine
the spin chain model in Eq. \eref{eq:heisenberg} equivalent to the system 
of trapped, strongly-interacting particles in Eq. \eref{eq:hamiltonian}. 

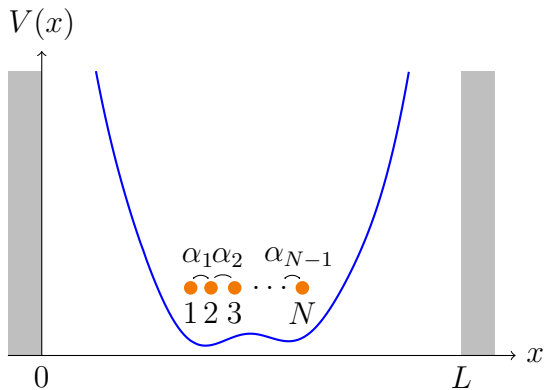
\begin{figure}[tbp]
  \centering
  \hspace*{-4mm}
  \begin{tikzpicture}[scale=.9]

    \begin{scope}[shift={(1,0)}];
      \fill [lightgray] (-.5,0) rectangle (0,4.2);
      \fill [lightgray] (6.2,0) rectangle (6.7,4.2);

      \draw [->] (-.5,0) -- (7,0);
      \node [below] at (0,0) {$0$};
      \node [below] at (6.2,0) {$L$};
      \node [right] at (7,0) {$x$};

      \draw [->] (0,0) -- (0,4.5);
      \node [above] at (0,4.5) {$V(x)$};

      \draw [blue, thick, domain=0.8:5.424, samples=100]
      plot (\x, {.15 + 10*abs(\x/3.1 - 1)^3 + .2*cos((\x + .56) r)*cos((3*\x) r)});
    \end{scope}
    \begin{scope}[shift={(0,.2)}]
      \node [below] at (3.2,.75) {$1$};
      \node [below] at (3.5,.75) {$2$};
      \node [below] at (3.85,.75) {$3$};
      \node [below] at (4.85,.75) {$N$};
      
      \fill [orange!95!black] (3.2,.8) circle (0.1);
      \fill [orange!95!black] (3.5,.8) circle (0.1);
      \fill [orange!95!black] (3.85,.8) circle (0.1);
      \node at (4.37,.8) {$\dots$};
      \fill [orange!95!black] (4.85,.8) circle (0.1);

      \draw [rounded corners] (3.25,.95) -- (3.35,1.05) -- (3.45,.95);
      \draw [rounded corners] (3.55,.95) -- (3.675,1.05) -- (3.8,.95);
      \draw [rounded corners] (4.6,.95) -- (4.7,1.05) -- (4.8,.95);

      \node [above] at (3.31,.95) {$\alpha_1$};
      \node [above] at (3.75,.95) {$\alpha_2$};
      \node [above] at (4.8,.93) {$\alpha_{N-1}$};
    \end{scope}
  \end{tikzpicture}
  \caption{Sketch of an $N$ particle spin chain in a trapping potential
    $V(x)$. We are able to calculate the spin-exchange coefficients 
    $\alpha_k$ governing the dynamics of the system for any potential 
    and $N \lesssim 35$. In our numerical implementation, the system 
    is placed in a hard-wall box ranging from $x=0$ to $x=L$. 
    This restricts the system to a finite region of space, but
    the length $L$ can be chosen so large that the cut-off does not 
    affect the $N$ lowest-energy single-particle states in the trap 
    potential $V(x)$ and hence the corresponding $\alpha_k$ coefficients.}
  \label{fig:spin-chain}
\end{figure}

Volosniev {\it et al.} \cite{volosniev2015} have derived the following
expression for the geometric coefficients,
\begin{equation}
  \label{eq:geometric}
  \alpha_k = 
  \int\limits_{x_1< x_2 <  \cdots < x_{N-1}}
  \textrm{d}x_1 \, \textrm{d}x_2 \cdots \textrm{d}x_{N-1}
  \left(
    \frac{\partial \Phi_0(x_1,\dots,x_N)}{\partial x_N}
  \right)^2_{x_N = x_k} .
\end{equation}
If we are interested in more than just a few particles, it quickly
becomes impossible to numerically evaluate the $N-1$ integrals of
Eq.~\eref{eq:geometric} accurately. Thus, in order to calculate the
geometric coefficients $\alpha_k$ for many-particle systems, the above
expression should be rewritten in a form that is better suited for numerical 
implementation. Such an expression was derived in Ref.~\cite{loft2016b}:
\begin{eqnarray}
  \label{eq:geometric-final}
  \alpha_k =
  &\sum_{i=1}^N \left[ \frac{\textrm{d}\psi_i}{\textrm{d}x}
  \right]^2_{x=b} +
  2 \sum_{i=1}^N \sum_{j=1}^N \sum_{l=0}^{N-1-k}
  \frac{(-1)^{i+j+N-k}}{l!}
  {N-l-2 \choose k-1} \nonumber\\
  &\times \int_a^b \textrm{d}x \,
  \frac{2m}{\hbar^2} \big( V(x) - E_i \big) \, \psi_i (x) \,
  \frac{\textrm{d}\psi_j}{\textrm{d}x}
  \left[
    \frac{\partial^l}{\partial\lambda^l}
    \det \big[ (B(x) - \lambda \textbf{I})^{(ij)} \big]
  \right]_{\lambda = 0} , \quad
\end{eqnarray}
where $B(x)$ is an $N \times N$ symmetric matrix constructed from
$\psi_i$, $i=1,\dots,N$, with $mn$'th entry given by $\int_a^x
\textrm{d}y \, \psi_m(y) \psi_n(y)$, and $(\;)^{(ij)}$ denotes a minor
obtained by removing the $i$'th column and the $j$'th row, while
$[a;b]$ is the spatial interval containing the system.  Despite its
complicated appearance, Eq.~\eref{eq:geometric-final} is amenable to
efficient numerical implementation, as detailed in
Ref.~\cite{loft2016b}, where we also present a well-working numerical
code in C freely available to anyone.  In our program, we place the
system in a box with impenetrable hard walls in order to fully confine
it to a finite region of space $[a;b] = [0;L]$, see
Figure~\ref{fig:spin-chain}.  The length of the box, $L$, can always
be chosen sufficiently large so as not to affect the $N$ lowest-energy
single-particle states in the trapping potential $V(x)$ that are
needed to compute the expression in Eq.~\eref{eq:geometric-final}. The
wavefunctions $\psi_i(x)$ are expanded in terms of the standing-wave
sinusoidal eigenfunctions for the box potential. Our program was first
used in Ref.~\cite{loft2015} to calculate the geometric coefficients
for $N \leq 30$ particles in the special case of a harmonic trap
potential. The calculation for $N = 30$ was accomplished in a matter
of hours on a small personal computer, while reliable results could be
easily obtained for even more particles.  The program can handle
arbitrary potentials, which we exploit in this paper to study the
quantum state transfer in engineered spin chains of nontrivial
lengths.

\section{Perfect state transfer potential}
\label{sec:perfect}

Consider a chain of $N$ interacting spin-$\frac{1}{2}$ fermions
described by the Heisenberg $XXZ$ model Hamiltonian
\begin{equation}
  \label{eq:sc-hamiltonian}
  H_\mathrm{SC} = - \frac{1}{2} \sum_{k=1}^{N-1} J_k
  \left( \sigma_x^k \sigma_x^{k+1} + \sigma_y^k \sigma_y^{k+1}
  + \Delta \, \sigma_z^k \sigma_z^{k+1} \right) \; ,
\end{equation}
where $J_k = -\alpha_k/g$ are the nearest-neighbor interaction coefficients, 
and $\Delta$ is the asymmetry parameter, which is related to 
the interaction parameter $\kappa$ via $\Delta = (1-2/\kappa)$. 
In the following, we consider the Heisenberg $XX$ model, 
i.e. we assume $\Delta = 0$ or $\kappa = 2$. 
Notice that $H_\mathrm{SC}$ conserves the total spin
projection $(N_\uparrow - N_\downarrow)/2$. Suppose that the initial 
state of the system at time $t=0$ is $\left| 1 \right> \equiv \left|\uparrow
  \downarrow \downarrow \cdots \downarrow \downarrow \right>$, meaning
that the particle at site 1 is in the spin-up state while the
remaining $N-1$ particles are spin-down. In general, this state is
not an eigenstate of the system, so the initial state will evolve with 
time into a linear combination of states with exactly one spin-up 
particle somewhere in the chain. We are interested in whether
the spin-up state is transferred to site $N$ at the opposite end of
the chain, i.e. the system ends up in state $\left| N \right> \equiv 
\left| \downarrow \downarrow \downarrow \cdots \downarrow \uparrow \right>$. 
The time dependent probability of successful spin transfer is 
\begin{equation}
  \label{eq:fidelity}
  F(t) = \left| \left< N \right| e^{-iH_\mathrm{SC}t/\hbar}
    \left| 1 \right>  \right|^2 \; .
\end{equation}
This quantity is known as the fidelity, and $F(t) = 1$ would correspond
to perfect transfer at time $t$.

It was shown in Refs. \cite{christandl2004,nikolopoulos2004,Plenio2004} 
that one can obtain perfect state or excitation transfer if the 
nearest-neighbor exchange coefficients $J_k \propto \alpha_k$ follow the form
\begin{equation}
  \label{eq:semicirc}
  \alpha_k \propto \sqrt{k(N-k)} \; ,
\end{equation}
corresponding to a semicircular shape. 
Remarkably, the optimal (fastest) and perfect state transfer is
obtained for any number $N$ of particles in the chain \cite{yung2006}. 
The pertinent question now is how to experimentally realize the 
system such that the interaction coefficients have the correct form. 
With a cold-atom implementation of the spin chain, this question 
reduces to that of choosing a trapping potential $V(x)$ 
of proper shape, which is what we investigate below.

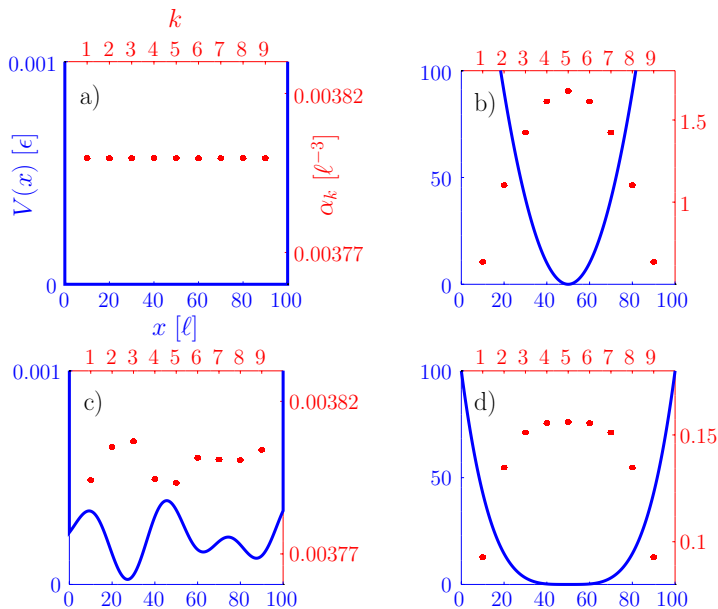
\begin{figure}[t]
  \centering
  \resizebox{.75\columnwidth}{!}{
    \setlength{\unitlength}{1pt}
    \begin{picture}(0,0)
      \includegraphics{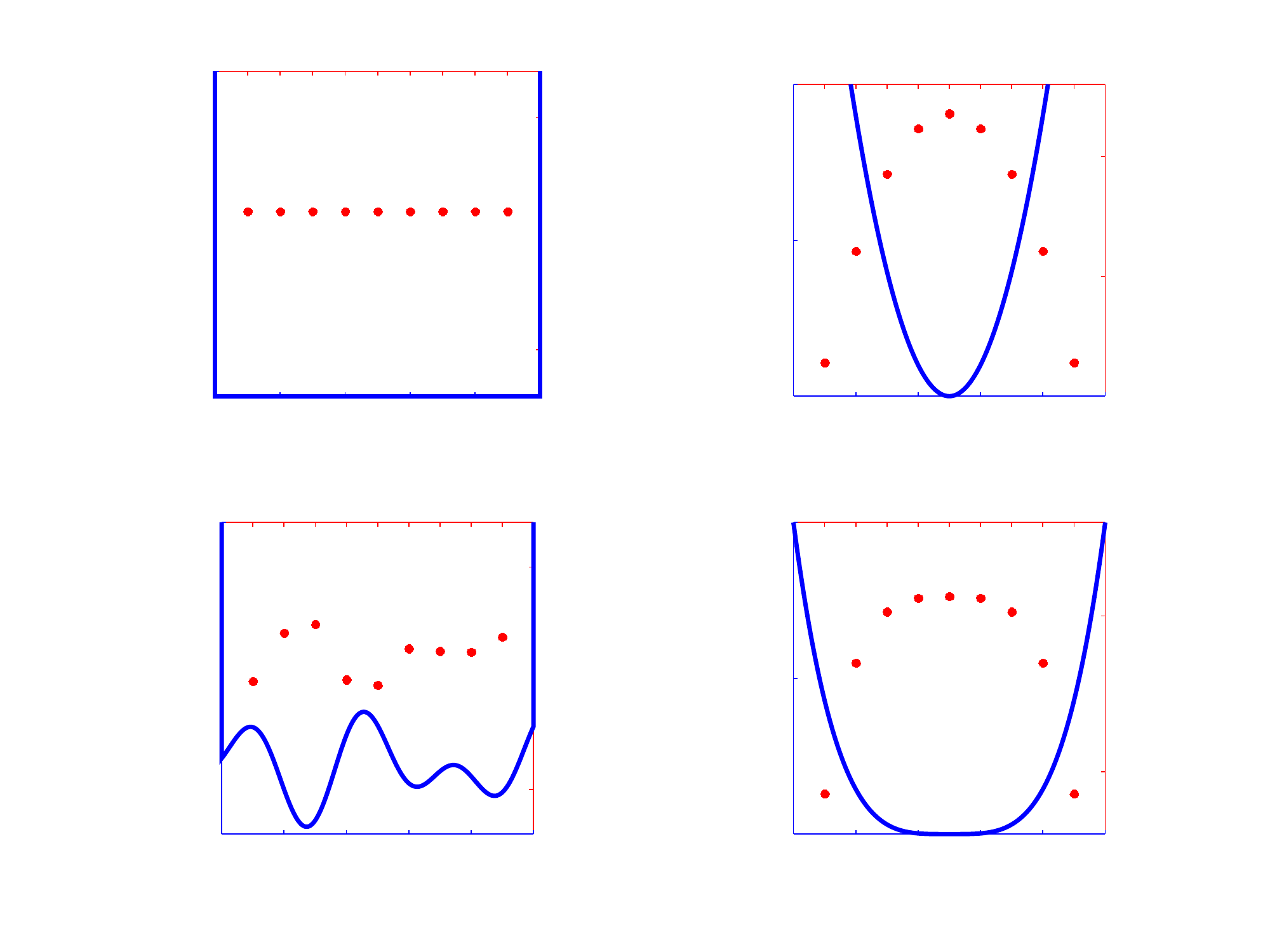}
    \end{picture}%
    \begin{picture}(576,432)(0,0)
      \fontsize{17}{0}
      \selectfont\put(171.262,424){\makebox(0,0)[b]{\textcolor[rgb]{1,0,0}{{$k$}}}}
      \fontsize{17}{0}
      \selectfont\put(171.262,215){\makebox(0,0)[b]{\textcolor[rgb]{0,0,1}{{$x$ $[\ell]$}}}}
      \fontsize{17}{0}
      \selectfont\put(80,325.909){\rotatebox{90}{\makebox(0,0)[b]{\textcolor[rgb]{0,0,1}{{$V(x)$ $[\epsilon]$}}}}}
      \fontsize{17}{0}
      \selectfont\put(280,325.909){\rotatebox{90}{\makebox(0,0)[b]{\textcolor[rgb]{1,0,0}{{$\alpha_k$ $[\ell^{-3}]$}}}}}
\fontsize{15}{0}
\selectfont\put(112.309,404.613){\makebox(0,0)[b]{\textcolor[rgb]{1,0,0}{{1}}}}
\fontsize{15}{0}
\selectfont\put(127.047,404.613){\makebox(0,0)[b]{\textcolor[rgb]{1,0,0}{{2}}}}
\fontsize{15}{0}
\selectfont\put(141.785,404.613){\makebox(0,0)[b]{\textcolor[rgb]{1,0,0}{{3}}}}
\fontsize{15}{0}
\selectfont\put(156.524,404.613){\makebox(0,0)[b]{\textcolor[rgb]{1,0,0}{{4}}}}
\fontsize{15}{0}
\selectfont\put(171.262,404.613){\makebox(0,0)[b]{\textcolor[rgb]{1,0,0}{{5}}}}
\fontsize{15}{0}
\selectfont\put(186,404.613){\makebox(0,0)[b]{\textcolor[rgb]{1,0,0}{{6}}}}
\fontsize{15}{0}
\selectfont\put(200.738,404.613){\makebox(0,0)[b]{\textcolor[rgb]{1,0,0}{{7}}}}
\fontsize{15}{0}
\selectfont\put(215.477,404.613){\makebox(0,0)[b]{\textcolor[rgb]{1,0,0}{{8}}}}
\fontsize{15}{0}
\selectfont\put(230.215,404.613){\makebox(0,0)[b]{\textcolor[rgb]{1,0,0}{{9}}}}
\fontsize{15}{0}
\selectfont\put(249.966,273.272){\makebox(0,0)[l]{\textcolor[rgb]{1,0,0}{{0.00377}}}}
\fontsize{15}{0}
\selectfont\put(249.966,378.545){\makebox(0,0)[l]{\textcolor[rgb]{1,0,0}{{0.00382}}}}
\fontsize{15}{0}
\selectfont\put(97.5707,247.205){\makebox(0,0)[t]{\textcolor[rgb]{0,0,1}{{0}}}}
\fontsize{15}{0}
\selectfont\put(127.047,247.205){\makebox(0,0)[t]{\textcolor[rgb]{0,0,1}{{20}}}}
\fontsize{15}{0}
\selectfont\put(156.524,247.205){\makebox(0,0)[t]{\textcolor[rgb]{0,0,1}{{40}}}}
\fontsize{15}{0}
\selectfont\put(186,247.205){\makebox(0,0)[t]{\textcolor[rgb]{0,0,1}{{60}}}}
\fontsize{15}{0}
\selectfont\put(215.477,247.205){\makebox(0,0)[t]{\textcolor[rgb]{0,0,1}{{80}}}}
\fontsize{15}{0}
\selectfont\put(244.953,247.205){\makebox(0,0)[t]{\textcolor[rgb]{0,0,1}{{100}}}}
\fontsize{15}{0}
\selectfont\put(92.5576,252.218){\makebox(0,0)[r]{\textcolor[rgb]{0,0,1}{{0}}}}
\fontsize{15}{0}
\selectfont\put(92.5576,399.6){\makebox(0,0)[r]{\textcolor[rgb]{0,0,1}{{0.001}}}}
\fontsize{17}{0}
\selectfont\put(107.887,377.493){\makebox(0,0)[l]{\textcolor[rgb]{0,0,0}{{a)}}}}
\fontsize{15}{0}
\selectfont\put(114.7,199.916){\makebox(0,0)[b]{\textcolor[rgb]{1,0,0}{{1}}}}
\fontsize{15}{0}
\selectfont\put(128.84,199.916){\makebox(0,0)[b]{\textcolor[rgb]{1,0,0}{{2}}}}
\fontsize{15}{0}
\selectfont\put(142.981,199.916){\makebox(0,0)[b]{\textcolor[rgb]{1,0,0}{{3}}}}
\fontsize{15}{0}
\selectfont\put(157.121,199.916){\makebox(0,0)[b]{\textcolor[rgb]{1,0,0}{{4}}}}
\fontsize{15}{0}
\selectfont\put(171.262,199.916){\makebox(0,0)[b]{\textcolor[rgb]{1,0,0}{{5}}}}
\fontsize{15}{0}
\selectfont\put(185.402,199.916){\makebox(0,0)[b]{\textcolor[rgb]{1,0,0}{{6}}}}
\fontsize{15}{0}
\selectfont\put(199.543,199.916){\makebox(0,0)[b]{\textcolor[rgb]{1,0,0}{{7}}}}
\fontsize{15}{0}
\selectfont\put(213.683,199.916){\makebox(0,0)[b]{\textcolor[rgb]{1,0,0}{{8}}}}
\fontsize{15}{0}
\selectfont\put(227.824,199.916){\makebox(0,0)[b]{\textcolor[rgb]{1,0,0}{{9}}}}
\fontsize{15}{0}
\selectfont\put(246.979,73.6983){\makebox(0,0)[l]{\textcolor[rgb]{1,0,0}{{0.00377}}}}
\fontsize{15}{0}
\selectfont\put(246.979,174.701){\makebox(0,0)[l]{\textcolor[rgb]{1,0,0}{{0.00382}}}}
\fontsize{15}{0}
\selectfont\put(100.56,48.4833){\makebox(0,0)[t]{\textcolor[rgb]{0,0,1}{{0}}}}
\fontsize{15}{0}
\selectfont\put(128.84,48.4833){\makebox(0,0)[t]{\textcolor[rgb]{0,0,1}{{20}}}}
\fontsize{15}{0}
\selectfont\put(157.121,48.4833){\makebox(0,0)[t]{\textcolor[rgb]{0,0,1}{{40}}}}
\fontsize{15}{0}
\selectfont\put(185.402,48.4833){\makebox(0,0)[t]{\textcolor[rgb]{0,0,1}{{60}}}}
\fontsize{15}{0}
\selectfont\put(213.683,48.4833){\makebox(0,0)[t]{\textcolor[rgb]{0,0,1}{{80}}}}
\fontsize{15}{0}
\selectfont\put(241.964,48.4833){\makebox(0,0)[t]{\textcolor[rgb]{0,0,1}{{100}}}}
\fontsize{15}{0}
\selectfont\put(95.5452,53.4977){\makebox(0,0)[r]{\textcolor[rgb]{0,0,1}{{0}}}}
\fontsize{15}{0}
\selectfont\put(95.5452,194.902){\makebox(0,0)[r]{\textcolor[rgb]{0,0,1}{{0.001}}}}
\fontsize{17}{0}
\selectfont\put(110.458,173.692){\makebox(0,0)[l]{\textcolor[rgb]{0,0,0}{{c)}}}}
\fontsize{15}{0}
\selectfont\put(373.998,398.637){\makebox(0,0)[b]{\textcolor[rgb]{1,0,0}{{1}}}}
\fontsize{15}{0}
\selectfont\put(388.139,398.637){\makebox(0,0)[b]{\textcolor[rgb]{1,0,0}{{2}}}}
\fontsize{15}{0}
\selectfont\put(402.279,398.637){\makebox(0,0)[b]{\textcolor[rgb]{1,0,0}{{3}}}}
\fontsize{15}{0}
\selectfont\put(416.42,398.637){\makebox(0,0)[b]{\textcolor[rgb]{1,0,0}{{4}}}}
\fontsize{15}{0}
\selectfont\put(430.56,398.637){\makebox(0,0)[b]{\textcolor[rgb]{1,0,0}{{5}}}}
\fontsize{15}{0}
\selectfont\put(444.7,398.637){\makebox(0,0)[b]{\textcolor[rgb]{1,0,0}{{6}}}}
\fontsize{15}{0}
\selectfont\put(458.841,398.637){\makebox(0,0)[b]{\textcolor[rgb]{1,0,0}{{7}}}}
\fontsize{15}{0}
\selectfont\put(472.981,398.637){\makebox(0,0)[b]{\textcolor[rgb]{1,0,0}{{8}}}}
\fontsize{15}{0}
\selectfont\put(487.122,398.637){\makebox(0,0)[b]{\textcolor[rgb]{1,0,0}{{9}}}}
\fontsize{15}{0}
\selectfont\put(506.277,306.604){\makebox(0,0)[l]{\textcolor[rgb]{1,0,0}{{1}}}}
\fontsize{15}{0}
\selectfont\put(506.277,360.99){\makebox(0,0)[l]{\textcolor[rgb]{1,0,0}{{1.5}}}}
\fontsize{15}{0}
\selectfont\put(359.858,247.203){\makebox(0,0)[t]{\textcolor[rgb]{0,0,1}{{0}}}}
\fontsize{15}{0}
\selectfont\put(388.139,247.203){\makebox(0,0)[t]{\textcolor[rgb]{0,0,1}{{20}}}}
\fontsize{15}{0}
\selectfont\put(416.42,247.203){\makebox(0,0)[t]{\textcolor[rgb]{0,0,1}{{40}}}}
\fontsize{15}{0}
\selectfont\put(444.7,247.203){\makebox(0,0)[t]{\textcolor[rgb]{0,0,1}{{60}}}}
\fontsize{15}{0}
\selectfont\put(472.981,247.203){\makebox(0,0)[t]{\textcolor[rgb]{0,0,1}{{80}}}}
\fontsize{15}{0}
\selectfont\put(501.262,247.203){\makebox(0,0)[t]{\textcolor[rgb]{0,0,1}{{100}}}}
\fontsize{15}{0}
\selectfont\put(354.843,252.218){\makebox(0,0)[r]{\textcolor[rgb]{0,0,1}{{0}}}}
\fontsize{15}{0}
\selectfont\put(354.843,322.92){\makebox(0,0)[r]{\textcolor[rgb]{0,0,1}{{50}}}}
\fontsize{15}{0}
\selectfont\put(354.843,393.622){\makebox(0,0)[r]{\textcolor[rgb]{0,0,1}{{100}}}}
\fontsize{17}{0}
\selectfont\put(369.756,372.412){\makebox(0,0)[l]{\textcolor[rgb]{0,0,0}{{b)}}}}
\fontsize{15}{0}
\selectfont\put(373.998,199.917){\makebox(0,0)[b]{\textcolor[rgb]{1,0,0}{{1}}}}
\fontsize{15}{0}
\selectfont\put(388.139,199.917){\makebox(0,0)[b]{\textcolor[rgb]{1,0,0}{{2}}}}
\fontsize{15}{0}
\selectfont\put(402.279,199.917){\makebox(0,0)[b]{\textcolor[rgb]{1,0,0}{{3}}}}
\fontsize{15}{0}
\selectfont\put(416.42,199.917){\makebox(0,0)[b]{\textcolor[rgb]{1,0,0}{{4}}}}
\fontsize{15}{0}
\selectfont\put(430.56,199.917){\makebox(0,0)[b]{\textcolor[rgb]{1,0,0}{{5}}}}
\fontsize{15}{0}
\selectfont\put(444.7,199.917){\makebox(0,0)[b]{\textcolor[rgb]{1,0,0}{{6}}}}
\fontsize{15}{0}
\selectfont\put(458.841,199.917){\makebox(0,0)[b]{\textcolor[rgb]{1,0,0}{{7}}}}
\fontsize{15}{0}
\selectfont\put(472.981,199.917){\makebox(0,0)[b]{\textcolor[rgb]{1,0,0}{{8}}}}
\fontsize{15}{0}
\selectfont\put(487.122,199.917){\makebox(0,0)[b]{\textcolor[rgb]{1,0,0}{{9}}}}
\fontsize{15}{0}
\selectfont\put(506.277,81.7786){\makebox(0,0)[l]{\textcolor[rgb]{1,0,0}{{0.1}}}}
\fontsize{15}{0}
\selectfont\put(506.277,152.481){\makebox(0,0)[l]{\textcolor[rgb]{1,0,0}{{0.15}}}}
\fontsize{15}{0}
\selectfont\put(359.858,48.4833){\makebox(0,0)[t]{\textcolor[rgb]{0,0,1}{{0}}}}
\fontsize{15}{0}
\selectfont\put(388.139,48.4833){\makebox(0,0)[t]{\textcolor[rgb]{0,0,1}{{20}}}}
\fontsize{15}{0}
\selectfont\put(416.42,48.4833){\makebox(0,0)[t]{\textcolor[rgb]{0,0,1}{{40}}}}
\fontsize{15}{0}
\selectfont\put(444.7,48.4833){\makebox(0,0)[t]{\textcolor[rgb]{0,0,1}{{60}}}}
\fontsize{15}{0}
\selectfont\put(472.981,48.4833){\makebox(0,0)[t]{\textcolor[rgb]{0,0,1}{{80}}}}
\fontsize{15}{0}
\selectfont\put(501.262,48.4833){\makebox(0,0)[t]{\textcolor[rgb]{0,0,1}{{100}}}}
\fontsize{15}{0}
\selectfont\put(354.843,53.4977){\makebox(0,0)[r]{\textcolor[rgb]{0,0,1}{{0}}}}
\fontsize{15}{0}
\selectfont\put(354.843,124.2){\makebox(0,0)[r]{\textcolor[rgb]{0,0,1}{{50}}}}
\fontsize{15}{0}
\selectfont\put(354.843,194.902){\makebox(0,0)[r]{\textcolor[rgb]{0,0,1}{{100}}}}
\fontsize{17}{0}
\selectfont\put(369.756,173.692){\makebox(0,0)[l]{\textcolor[rgb]{0,0,0}{{d)}}}}
\end{picture}
  }
  \caption{The amplitudes of geometric coefficients approximately follow 
    the shape of $-V(x)$.  For $N=10$ particles, we show the potential $V(x)$
    (blue line) and the geometric coefficients $\alpha_k$ (red dots)
    for (a) hard wall potential, (b) harmonic potential, (c) hard wall
    potential with noise and (d) our optimized state transfer
    potential of Eq.~\eref{eq:semicirc-potential} with $\tau =
    3.793287$. The units are indicated on the corresponding axes of
    (a).}
  \label{fig:potential-alpha}
\end{figure}

\subsection{Connection between $V(x)$ and the $\alpha_k$'s}
Our goal is to determine the shape of the trapping potential $V(x)$ 
which results in the geometric coefficients following the distribution 
of Eq. \eref{eq:semicirc} for an $N$-particle spin chain. 
With the length scale of the potential $\ell$, the corresponding energy
scale is $\epsilon = \hbar^2/(2m\ell^2)$. In our numerical program, 
we place the system into a hard-wall box of large size $L=100\ell$.

Before describing a class of candidate potentials, we present a useful 
observation concerning the relation between the shapes of $V(x)$ and 
$\alpha_k$'s. The geometric coefficients of Eq.~\eref{eq:geometric} 
only depend on $V(x)$ through the Slater determinant wavefunction, 
$\Phi_0$, composed of the eigenstates of the single-particle 
Hamiltonian \eref{eq:single-particle-hamiltonian}. Even though we 
start with a simple function $V(x)$ and end up with a set of $N-1$ 
numbers $\alpha_k$, the precise relation between the two is still extremely 
complicated. However, as is evident from Figure~\ref{fig:potential-alpha},
the geometric coefficients approximately follow the shape of $-V(x)$.
An intuitive explanation is that the deeper is the potential, 
the more is the overlap of the spatial wavefunctions of neighboring particles, 
and therefore the larger is the corresponding geometric coefficient describing 
the spin-exchange amplitude. This seems to imply that a local density 
approximation should work well in some cases. For a detailed study of this 
question we refer the reader to Ref.~\cite{oleksandr2015}.

\begin{figure}[t]
  \centering
  \resizebox{.75\columnwidth}{!}{
\setlength{\unitlength}{1pt}
\begin{picture}(0,0)
\includegraphics{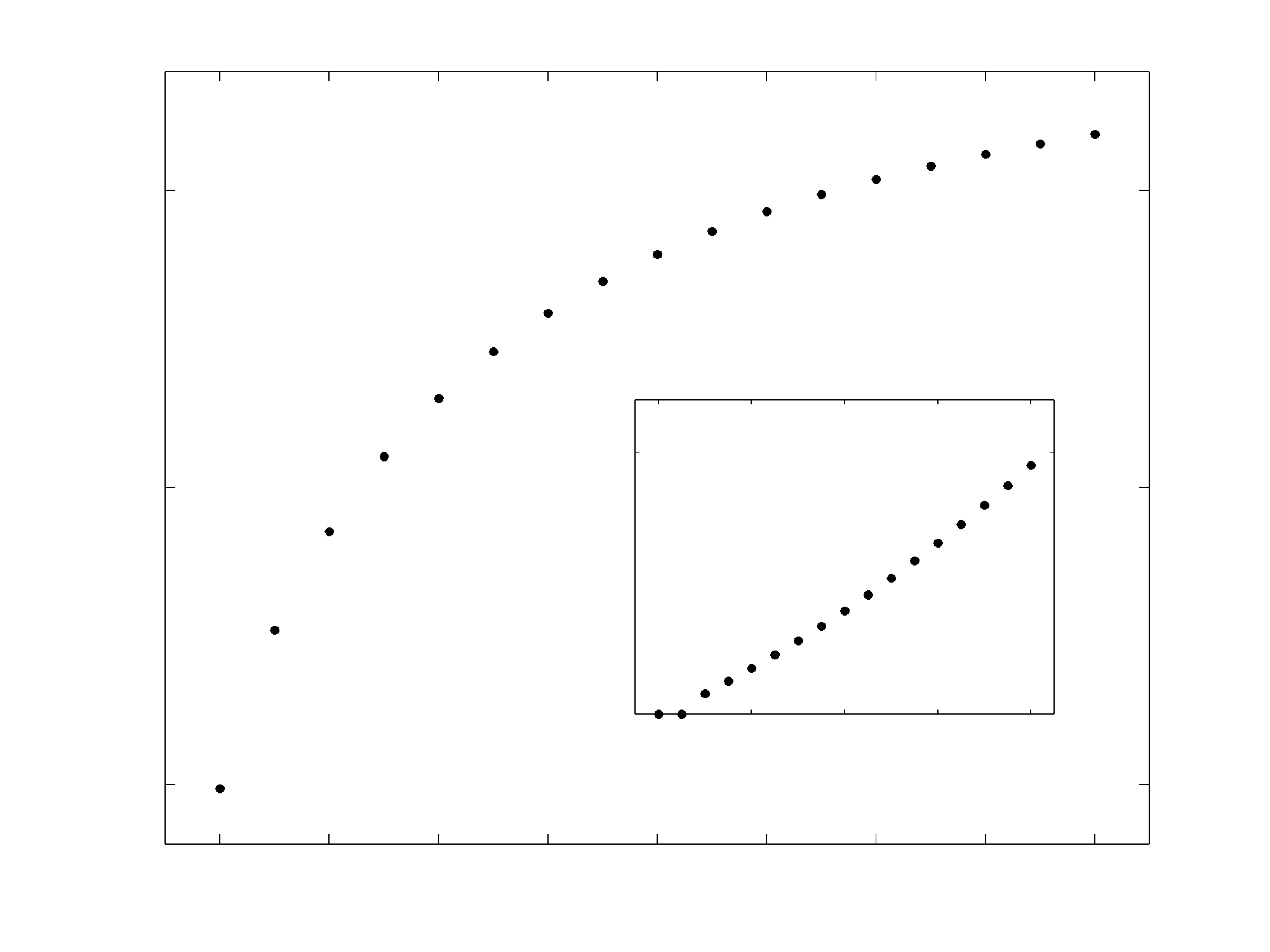}
\end{picture}%
\begin{picture}(576,432)(0,0)
\fontsize{18}{0}
\selectfont\put(99.68,44.0085){\makebox(0,0)[t]{\textcolor[rgb]{0,0,0}{{4}}}}
\fontsize{18}{0}
\selectfont\put(149.28,44.0085){\makebox(0,0)[t]{\textcolor[rgb]{0,0,0}{{6}}}}
\fontsize{18}{0}
\selectfont\put(198.88,44.0085){\makebox(0,0)[t]{\textcolor[rgb]{0,0,0}{{8}}}}
\fontsize{18}{0}
\selectfont\put(248.48,44.0085){\makebox(0,0)[t]{\textcolor[rgb]{0,0,0}{{10}}}}
\fontsize{18}{0}
\selectfont\put(298.08,44.0085){\makebox(0,0)[t]{\textcolor[rgb]{0,0,0}{{12}}}}
\fontsize{18}{0}
\selectfont\put(347.68,44.0085){\makebox(0,0)[t]{\textcolor[rgb]{0,0,0}{{14}}}}
\fontsize{18}{0}
\selectfont\put(397.28,44.0085){\makebox(0,0)[t]{\textcolor[rgb]{0,0,0}{{16}}}}
\fontsize{18}{0}
\selectfont\put(446.88,44.0085){\makebox(0,0)[t]{\textcolor[rgb]{0,0,0}{{18}}}}
\fontsize{18}{0}
\selectfont\put(496.48,44.0085){\makebox(0,0)[t]{\textcolor[rgb]{0,0,0}{{20}}}}
\fontsize{18}{0}
\selectfont\put(69.8755,75.9718){\makebox(0,0)[r]{\textcolor[rgb]{0,0,0}{{3}}}}
\fontsize{18}{0}
\selectfont\put(69.8755,210.817){\makebox(0,0)[r]{\textcolor[rgb]{0,0,0}{{3.5}}}}
\fontsize{18}{0}
\selectfont\put(69.8755,345.662){\makebox(0,0)[r]{\textcolor[rgb]{0,0,0}{{4}}}}
\fontsize{18}{0}
\selectfont\put(298.08,22.0086){\makebox(0,0)[t]{\textcolor[rgb]{0,0,0}{{$N$}}}}
\fontsize{18}{0}
\selectfont\put(37.8755,224.301){\rotatebox{90}{\makebox(0,0)[b]{\textcolor[rgb]{0,0,0}{{$\tau$}}}}}
\fontsize{18}{0}
\selectfont\put(298.56,103.015){\makebox(0,0)[t]{\textcolor[rgb]{0,0,0}{{4}}}}
\fontsize{18}{0}
\selectfont\put(340.8,103.015){\makebox(0,0)[t]{\textcolor[rgb]{0,0,0}{{8}}}}
\fontsize{18}{0}
\selectfont\put(383.04,103.015){\makebox(0,0)[t]{\textcolor[rgb]{0,0,0}{{12}}}}
\fontsize{18}{0}
\selectfont\put(425.28,103.015){\makebox(0,0)[t]{\textcolor[rgb]{0,0,0}{{16}}}}
\fontsize{18}{0}
\selectfont\put(467.52,103.015){\makebox(0,0)[t]{\textcolor[rgb]{0,0,0}{{20}}}}
\fontsize{18}{0}
\selectfont\put(282.998,108){\makebox(0,0)[r]{\textcolor[rgb]{0,0,0}{{0}}}}
\fontsize{18}{0}
\selectfont\put(282.998,226.8){\makebox(0,0)[r]{\textcolor[rgb]{0,0,0}{{0.005}}}}
\fontsize{18}{0}
\selectfont\put(383.04,81.0154){\makebox(0,0)[t]{\textcolor[rgb]{0,0,0}{{$N$}}}}
\fontsize{18}{0}
\selectfont\put(255,179.28){\rotatebox{90}{\makebox(0,0)[b]{\textcolor[rgb]{0,0,0}{{$f$}}}}}
\end{picture}
  }
  \caption{For the $\tau$-dependent trial potential
    \eref{eq:semicirc-potential} we fit the corresponding geometric
    coefficients to Eq.~\eref{eq:alpha-fit}. The main panel shows values
    of $\tau$ optimized to produce the fit parameter $\beta = 1/2$ for
    different particle numbers $N$. The inset shows the corresponding
    goodness of the fit $f$.}
  \label{fig:optim}
\end{figure}
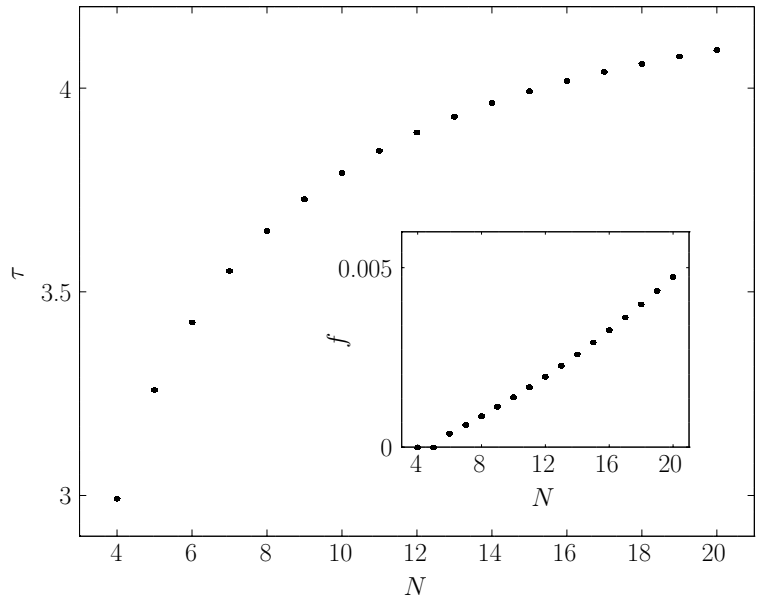

This observation motivates us to choose a trial trapping potential 
which resembles the shape of semicircle. We choose the simple ``bowl-shaped'' 
potential (see Figure~\ref{fig:potential-alpha}(d)).
\begin{equation}
  \label{eq:semicirc-potential}
  V(x) = 100\epsilon \cdot \left| \frac{L/2 - x}{L/2} \right|^\tau \; ,
\end{equation}
where $\tau$ is an adjustable parameter. The scale factor of $100$ 
is set large enough to ensure that the hard-wall boundaries do not affect 
the $N$ lowest-energy single-particle states in $V(x)$ and thereby the
calculation of the geometric coefficients
\footnote{Notice that $V(0) = V(L) = 100\epsilon$, whereas the highest 
single-particle state we use has much smaller energy of $7.80 \epsilon$ 
(the energy of the 20th state for $N=20$), so the hard wall boundaries 
do not influence the calculation of the geometric coefficients.}.
The idea now is quite simple: For any given $N$, we calculate 
the set of geometric coefficients corresponding to the potential in
Eq.~\eref{eq:semicirc-potential} for some $\tau$. We then fit the
produced $N-1$ geometric coefficients to a function of the form
\begin{equation}
  \label{eq:alpha-fit}
  \alpha_k \propto \big[ k (N-k) \big]^\beta \; ,
\end{equation}
where $\beta$ is a fit parameter such that the desired semicircle
distribution of Eq.~\eref{eq:semicirc} corresponds to $\beta = 1/2$. 
If we can find a value of $\tau$ for which $\beta = 1/2$, and if the 
fit is good, we expect the system to exhibit perfect state transfer. 
Since the trial potential of Eq.~\eref{eq:semicirc-potential} is 
rather simple, with only one adjustable parameter, we do not in general 
expect to be able to obtain a perfect fit of the geometric coefficients. 
Optimizing $\tau$ such that the best fit produces $\beta = 1/2$ yields 
the results of Figure~\ref{fig:optim}, where we also show a measure 
for the goodness of the fit 
\footnote{Let $(y_i,x_i)$ be the data points for $i=1,\dots,n$ we
  wish to fit to the model function $y(x)$. We define the goodness 
  of the fit as the length of the vector of differences 
  $f_i = y(x_i) - y_i$ normalized by the number of data points,
  $f = \frac{1}{n} \sqrt{\sum_{i=1}^n f_i^2}$; 
  clearly smaller $f$ corresponds to better fit.}.
It is not surprising that the fit is better for smaller $N$ because 
the number of fit points is $N-1$. The geometric coefficients corresponding 
to the optimized potential are given in \ref{sec:appendix}.

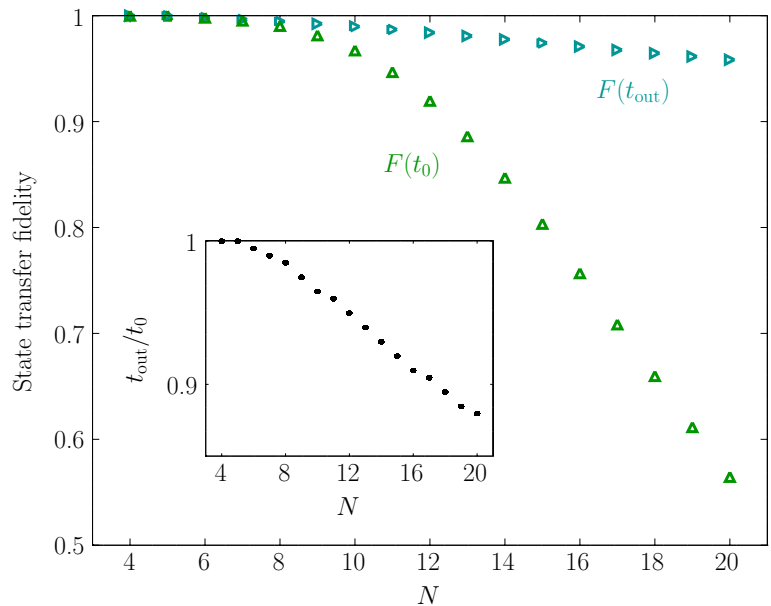
\begin{figure}[tbp]
  \centering
  \resizebox{.75\columnwidth}{!}{
    \setlength{\unitlength}{1pt}
\begin{picture}(0,0)
\includegraphics{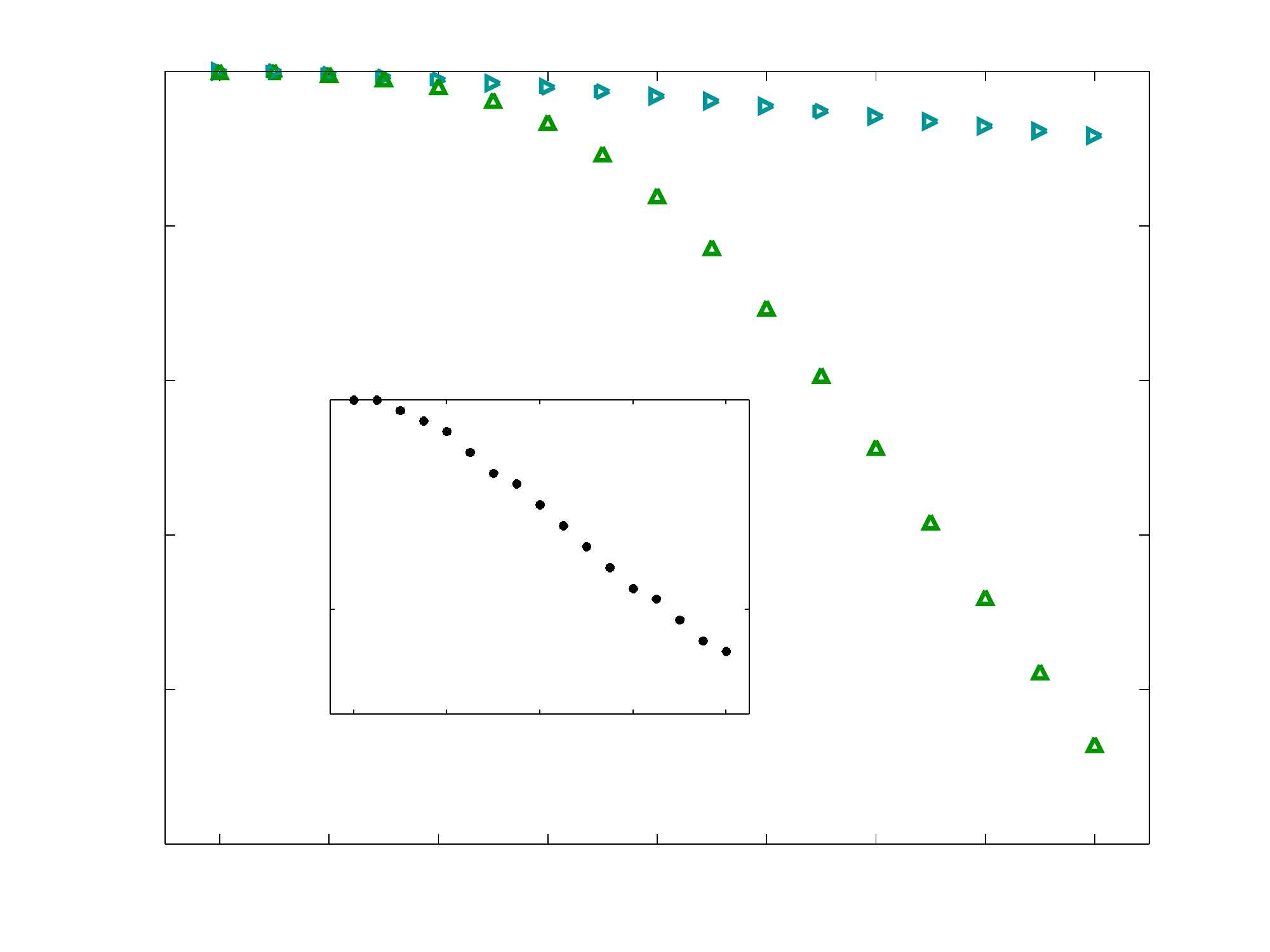}
\end{picture}%
\begin{picture}(576,432)(0,0)
\fontsize{18}{0}
\selectfont\put(99.68,44.0085){\makebox(0,0)[t]{\textcolor[rgb]{0,0,0}{{4}}}}
\fontsize{18}{0}
\selectfont\put(149.28,44.0085){\makebox(0,0)[t]{\textcolor[rgb]{0,0,0}{{6}}}}
\fontsize{18}{0}
\selectfont\put(198.88,44.0085){\makebox(0,0)[t]{\textcolor[rgb]{0,0,0}{{8}}}}
\fontsize{18}{0}
\selectfont\put(248.48,44.0085){\makebox(0,0)[t]{\textcolor[rgb]{0,0,0}{{10}}}}
\fontsize{18}{0}
\selectfont\put(298.08,44.0085){\makebox(0,0)[t]{\textcolor[rgb]{0,0,0}{{12}}}}
\fontsize{18}{0}
\selectfont\put(347.68,44.0085){\makebox(0,0)[t]{\textcolor[rgb]{0,0,0}{{14}}}}
\fontsize{18}{0}
\selectfont\put(397.28,44.0085){\makebox(0,0)[t]{\textcolor[rgb]{0,0,0}{{16}}}}
\fontsize{18}{0}
\selectfont\put(446.88,44.0085){\makebox(0,0)[t]{\textcolor[rgb]{0,0,0}{{18}}}}
\fontsize{18}{0}
\selectfont\put(496.48,44.0085){\makebox(0,0)[t]{\textcolor[rgb]{0,0,0}{{20}}}}
\fontsize{18}{0}
\selectfont\put(69.8755,49.0028){\makebox(0,0)[r]{\textcolor[rgb]{0,0,0}{{0.5}}}}
\fontsize{18}{0}
\selectfont\put(69.8755,119.122){\makebox(0,0)[r]{\textcolor[rgb]{0,0,0}{{0.6}}}}
\fontsize{18}{0}
\selectfont\put(69.8755,189.242){\makebox(0,0)[r]{\textcolor[rgb]{0,0,0}{{0.7}}}}
\fontsize{18}{0}
\selectfont\put(69.8755,259.361){\makebox(0,0)[r]{\textcolor[rgb]{0,0,0}{{0.8}}}}
\fontsize{18}{0}
\selectfont\put(69.8755,329.481){\makebox(0,0)[r]{\textcolor[rgb]{0,0,0}{{0.9}}}}
\fontsize{18}{0}
\selectfont\put(69.8755,399.6){\makebox(0,0)[r]{\textcolor[rgb]{0,0,0}{{1}}}}
\fontsize{18}{0}
\selectfont\put(298.08,22.0085){\makebox(0,0)[t]{\textcolor[rgb]{0,0,0}{{$N$}}}}
\fontsize{18}{0}
\selectfont\put(37.8755,224.301){\rotatebox{90}{\makebox(0,0)[b]{\textcolor[rgb]{0,0,0}{{State transfer fidelity}}}}}
\fontsize{18}{0}
\selectfont\put(409.68,350.516){\makebox(0,0)[l]{\textcolor[rgb]{0,0.588235,0.588235}{{$F(t_\text{out})$}}}}
\fontsize{18}{0}
\selectfont\put(268.32,301.433){\makebox(0,0)[l]{\textcolor[rgb]{0,0.588235,0}{{$F(t_0)$}}}}
\fontsize{18}{0}
\selectfont\put(160.32,103.015){\makebox(0,0)[t]{\textcolor[rgb]{0,0,0}{{4}}}}
\fontsize{18}{0}
\selectfont\put(202.56,103.015){\makebox(0,0)[t]{\textcolor[rgb]{0,0,0}{{8}}}}
\fontsize{18}{0}
\selectfont\put(244.8,103.015){\makebox(0,0)[t]{\textcolor[rgb]{0,0,0}{{12}}}}
\fontsize{18}{0}
\selectfont\put(287.04,103.015){\makebox(0,0)[t]{\textcolor[rgb]{0,0,0}{{16}}}}
\fontsize{18}{0}
\selectfont\put(329.28,103.015){\makebox(0,0)[t]{\textcolor[rgb]{0,0,0}{{20}}}}
\fontsize{18}{0}
\selectfont\put(144.758,155.52){\makebox(0,0)[r]{\textcolor[rgb]{0,0,0}{{0.9}}}}
\fontsize{18}{0}
\selectfont\put(144.758,250.56){\makebox(0,0)[r]{\textcolor[rgb]{0,0,0}{{1}}}}
\fontsize{18}{0}
\selectfont\put(244.8,81.0154){\makebox(0,0)[t]{\textcolor[rgb]{0,0,0}{{$N$}}}}
\fontsize{18}{0}
\selectfont\put(112.758,179.28){\rotatebox{90}{\makebox(0,0)[b]{\textcolor[rgb]{0,0,0}{{$t_\text{out}/t_0$}}}}}
\end{picture}
  }
  \caption{The main panel shows the transfer fidelity $F$ at times $t_0$ and $t_\mathrm{out}$
    for different $N$. The inset shows the corresponding ratio $t_\mathrm{out}/t_0$.}
  \label{fig:fidoptim}
\end{figure}

\subsection{Perfect state transfer}
To check whether the optimized potentials lead to perfect state transfer, 
we calculate the fidelity of Eq.~\eref{eq:fidelity}. 
In the Heisenberg $XX$ spin chain with coupling constants given by 
$J_k \propto -\alpha_k \propto -\sqrt{k(N-k)}$, the optimal transfer 
time is $t_0 = \hbar \pi\sqrt{N-1}/2|J_1|$ 
\cite{christandl2004,nikolopoulos2004,yung2006}. 
In Figure~\ref{fig:fidoptim} we observe that the fidelity at $t=t_0$ does 
indeed indicate a nearly-perfect transfer for the smaller values of $N$ 
we consider, but then $F(t_0)$ quickly decreases for $N \geq 10$.
Generally, the maximum of $F(t)$ occurs at a time $t_\mathrm{out}$ slightly 
before $t_0$, see the inset of Figure~\ref{fig:fidoptim}. We may think 
of $t_\mathrm{out}$ as the retrieval (or measurement) time, and we obtain
good transfer fidelity $F(t_\mathrm{out}) \geq 0.95$ for all $N \leq 20$. 
Next, we will study how the fidelity is affected by noise stemming 
from fluctuations of the potential.

\section{Noise tolerant state transfer}
\label{sec:noise}
In the previous section, we saw that the atoms in a trapping potential 
of the form of Eq.~\eref{eq:semicirc-potential} form a self-assembled 
spin chain which can mediate nearly-perfect state transfer between 
its two ends. Our theoretical calculations so far assumed 
idealized situation in which the potential is well-defined 
and engineered to maximize the transfer fidelity. In any real experiment,
however, there will be uncontrollable fluctuations of the potential. 
In this section we study how the resulting noise will influence the 
retrieval fidelity $F(t_\textrm{out})$. We emphasize that we consider 
uncontrolled variations of the potential leading to a noisy spin chain. 
In other words, we treat the noise exactly by including it in 
the original system, before we derive an effective spin chain
parameters. This is in contrast to the usual approach 
\cite{DeChiara2005,petrosyan2010,Zwick2011,Nikolopoulos2013}, 
in which one typically includes uniform, Gaussian or colored noise 
in the interaction parameters of the spin lattice model, without
specifying the original source of the fluctuations. 

\subsection{Random noise on the potential}
\label{sec:random-pot}
We add a noise term $\delta V(x)$ to the optimized potential 
$V(x)$ of Eq.~\eref{eq:semicirc-potential} in the form of a
quasi-periodic potential \cite{diener2001}:
\begin{equation}
  \label{eq:Vnoise}
  \delta V(x) = V_0 \left[ \cos\left(x/\ell + \phi_1\right)
  + \cos\left(\xi x /\ell + \phi_2\right) \right] \; ,
\end{equation}
where $V_0$ is the strength of the noise, $\phi_1, \phi_2 \in
[0;2\pi[$ are uniformly distributed random phases, and $\xi$ is an
irrational number chosen to be $\xi = 2/(1+\sqrt{5}) \approx 0.618$. 

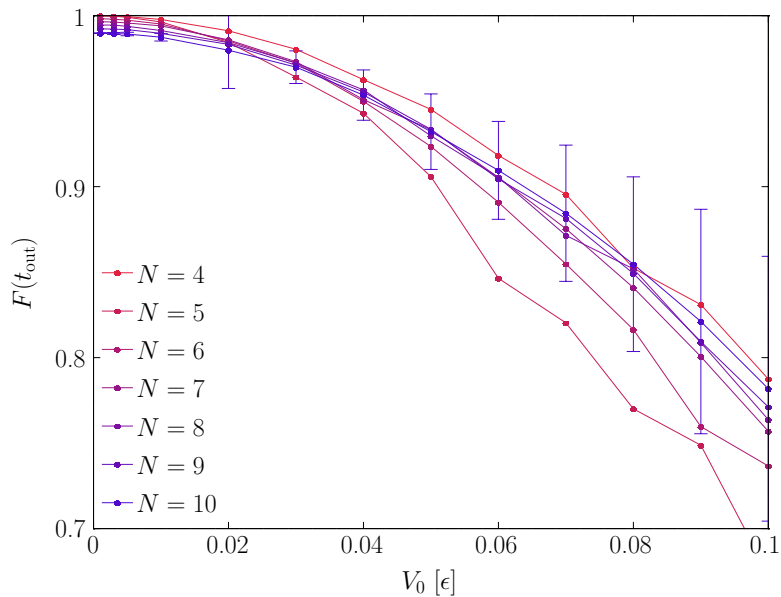
\begin{figure}[tbp]
  \centering
  \resizebox{.75\columnwidth}{!}{
\setlength{\unitlength}{1pt}
\begin{picture}(0,0)
\includegraphics{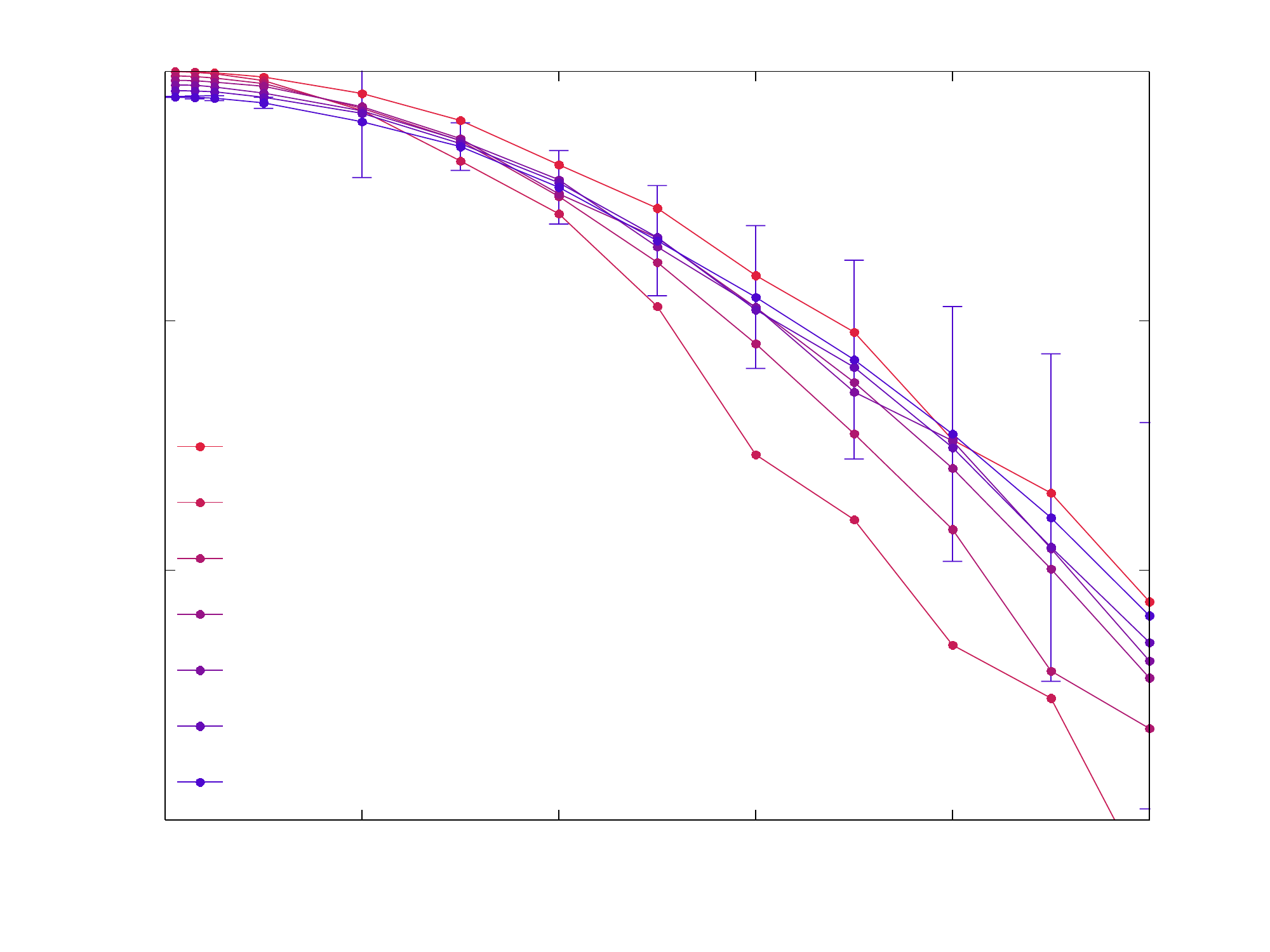}
\end{picture}%
\begin{picture}(576,432)(0,0)
\fontsize{18}{0}
\selectfont\put(74.88,55.004){\makebox(0,0)[t]{\textcolor[rgb]{0,0,0}{{0}}}}
\fontsize{18}{0}
\selectfont\put(164.16,55.004){\makebox(0,0)[t]{\textcolor[rgb]{0,0,0}{{0.02}}}}
\fontsize{18}{0}
\selectfont\put(253.44,55.004){\makebox(0,0)[t]{\textcolor[rgb]{0,0,0}{{0.04}}}}
\fontsize{18}{0}
\selectfont\put(342.72,55.004){\makebox(0,0)[t]{\textcolor[rgb]{0,0,0}{{0.06}}}}
\fontsize{18}{0}
\selectfont\put(432,55.004){\makebox(0,0)[t]{\textcolor[rgb]{0,0,0}{{0.08}}}}
\fontsize{18}{0}
\selectfont\put(521.28,55.004){\makebox(0,0)[t]{\textcolor[rgb]{0,0,0}{{0.1}}}}
\fontsize{18}{0}
\selectfont\put(69.8755,59.9982){\makebox(0,0)[r]{\textcolor[rgb]{0,0,0}{{0.7}}}}
\fontsize{18}{0}
\selectfont\put(69.8755,173.199){\makebox(0,0)[r]{\textcolor[rgb]{0,0,0}{{0.8}}}}
\fontsize{18}{0}
\selectfont\put(69.8755,286.399){\makebox(0,0)[r]{\textcolor[rgb]{0,0,0}{{0.9}}}}
\fontsize{18}{0}
\selectfont\put(69.8755,399.6){\makebox(0,0)[r]{\textcolor[rgb]{0,0,0}{{1}}}}
\fontsize{18}{0}
\selectfont\put(298.08,33.004){\makebox(0,0)[t]{\textcolor[rgb]{0,0,0}{{$V_0$ $[\epsilon]$}}}}
\fontsize{18}{0}
\selectfont\put(37.8755,229.799){\rotatebox{90}{\makebox(0,0)[b]{\textcolor[rgb]{0,0,0}{{$F(t_\text{out})$}}}}}
\fontsize{18}{0}
\selectfont\put(103.703,229.395){\makebox(0,0)[l]{\textcolor[rgb]{0,0,0}{{$N=4$}}}}
\fontsize{18}{0}
\selectfont\put(103.703,204.022){\makebox(0,0)[l]{\textcolor[rgb]{0,0,0}{{$N=5$}}}}
\fontsize{18}{0}
\selectfont\put(103.703,178.648){\makebox(0,0)[l]{\textcolor[rgb]{0,0,0}{{$N=6$}}}}
\fontsize{18}{0}
\selectfont\put(103.703,153.275){\makebox(0,0)[l]{\textcolor[rgb]{0,0,0}{{$N=7$}}}}
\fontsize{18}{0}
\selectfont\put(103.703,127.901){\makebox(0,0)[l]{\textcolor[rgb]{0,0,0}{{$N=8$}}}}
\fontsize{18}{0}
\selectfont\put(103.703,102.527){\makebox(0,0)[l]{\textcolor[rgb]{0,0,0}{{$N=9$}}}}
\fontsize{18}{0}
\selectfont\put(103.703,77.1538){\makebox(0,0)[l]{\textcolor[rgb]{0,0,0}{{$N=10$}}}}
\end{picture}
  }
  \caption{Retrieval fidelity, $F(t_\text{out})$, as a function of
    noise strength $V_0$, for different particle numbers $N$. The
    error bars for $N=10$ are one standard deviation. The standard
    deviations for other values of $N$ are similar, but not shown for
    better readability.}
  \label{fig:noise}
\end{figure}

To quantify how $F(t_\text{out})$, for a given number of particles $N
\leq 10$, is affected by the noise of Eq.~\eref{eq:Vnoise}, we take an
ensemble of $M=200$ random phases $\phi_1$ and $\phi_2$ for each noise
strength $V_0$ and calculate the corresponding $M$ sets of geometric
coefficients. The resulting average fidelities $F(t_\text{out})$ 
and their uncertainties (for the case of $N=10$) are shown in
Figure~\ref{fig:noise}.  The noise strength $V_0$ should be compared
to the typical energy scale of the system $\epsilon$ \footnote{For $N
  = 4,\dots,10$, the energy of the $N$'th single-particle state ranges
  from $0.55\epsilon$ to $1.21\epsilon$, which characterizes the
  energy scale of the system.}.  We thus verify that the state transfer 
can tolerate uncontrolled variations of the trapping potential at a few 
percent level, $V_0 < 0.02\epsilon$. We estimate that for a potential $V(x)$ 
with a typical length scale $\ell$ of several microns, the energy scale 
$\epsilon \approx 500\:$Hz. This then leads to $V_0 < 10\:$Hz, 
which is experimentally realistic but not trivial.

In Figure~\ref{fig:noise} we observe that for a weak noise the fidelity 
is larger for smaller particle numbers $N$, which is consistent with 
our observation above that in the noise-free limit $F(t_\text{out})$ 
slowly decreases with increasing $N$. With increasing the noise strengths,
however, the fidelity generally decreases faster for smaller values of 
$N$ (except for $N=4$), which means that longer chains are more robust.
A probable reason for this behavior is averaging out of the potential
fluctuations which is perhaps more complete for larger systems.

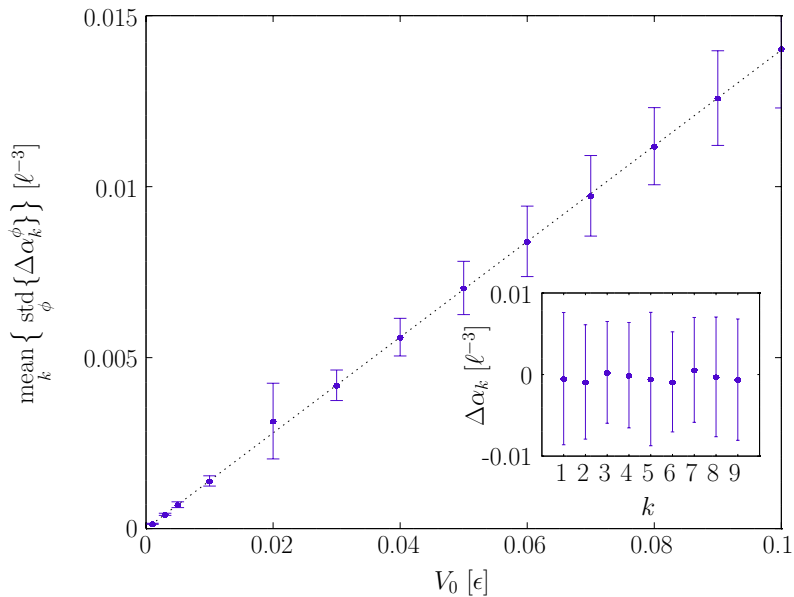
\begin{figure}[tbp]
  \centering
  \resizebox{.75\columnwidth}{!}{
    \setlength{\unitlength}{1pt}
\begin{picture}(0,0)
\includegraphics{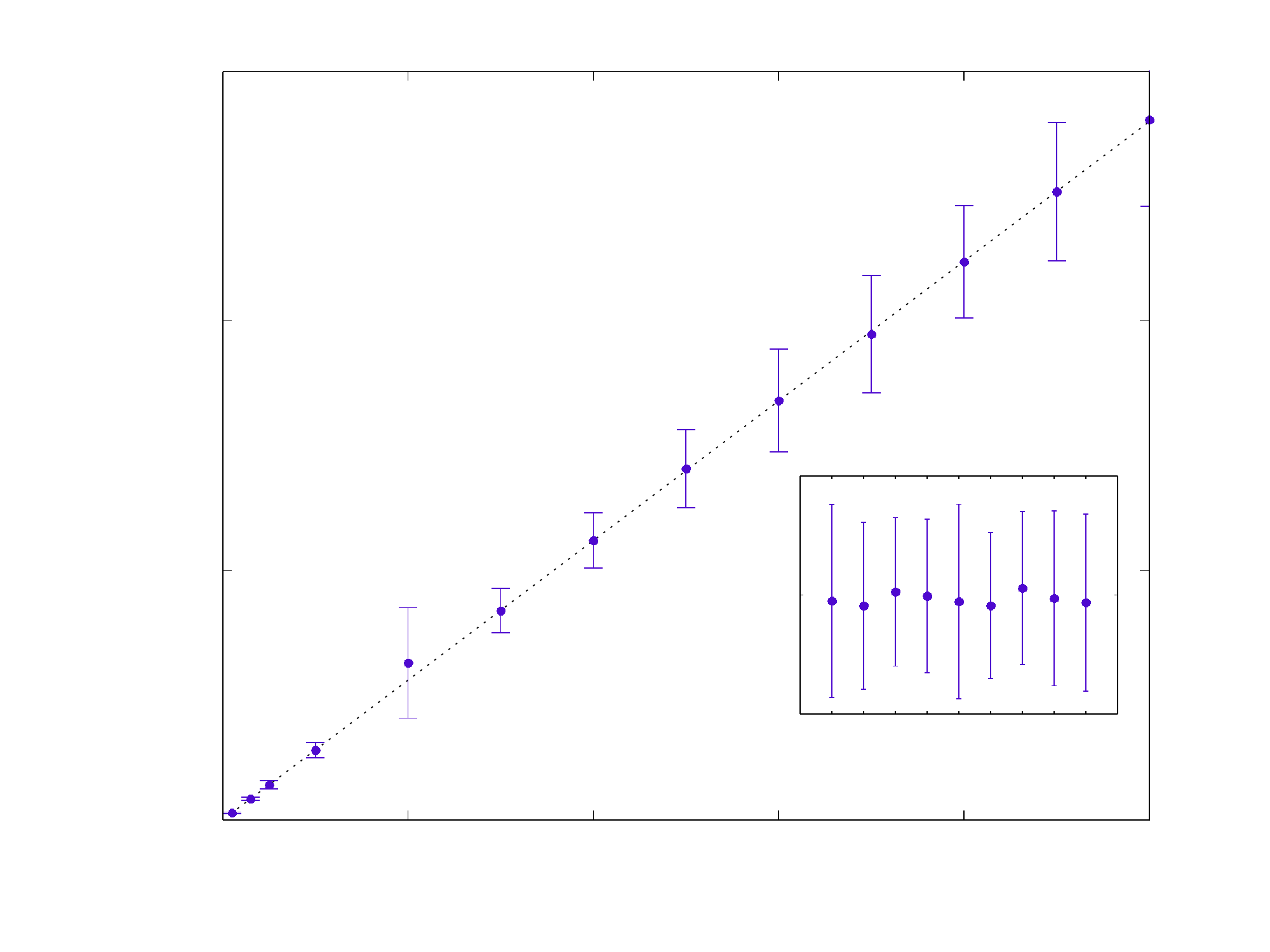}
\end{picture}%
\begin{picture}(576,432)(0,0)
\fontsize{18}{0}
\selectfont\put(101.005,55.0041){\makebox(0,0)[t]{\textcolor[rgb]{0,0,0}{{0}}}}
\fontsize{18}{0}
\selectfont\put(185.06,55.0041){\makebox(0,0)[t]{\textcolor[rgb]{0,0,0}{{0.02}}}}
\fontsize{18}{0}
\selectfont\put(269.115,55.0041){\makebox(0,0)[t]{\textcolor[rgb]{0,0,0}{{0.04}}}}
\fontsize{18}{0}
\selectfont\put(353.17,55.0041){\makebox(0,0)[t]{\textcolor[rgb]{0,0,0}{{0.06}}}}
\fontsize{18}{0}
\selectfont\put(437.225,55.0041){\makebox(0,0)[t]{\textcolor[rgb]{0,0,0}{{0.08}}}}
\fontsize{18}{0}
\selectfont\put(521.28,55.0041){\makebox(0,0)[t]{\textcolor[rgb]{0,0,0}{{0.1}}}}
\fontsize{18}{0}
\selectfont\put(96.0014,59.9982){\makebox(0,0)[r]{\textcolor[rgb]{0,0,0}{{0}}}}
\fontsize{18}{0}
\selectfont\put(96.0014,173.199){\makebox(0,0)[r]{\textcolor[rgb]{0,0,0}{{0.005}}}}
\fontsize{18}{0}
\selectfont\put(96.0014,286.399){\makebox(0,0)[r]{\textcolor[rgb]{0,0,0}{{0.01}}}}
\fontsize{18}{0}
\selectfont\put(96.0014,399.6){\makebox(0,0)[r]{\textcolor[rgb]{0,0,0}{{0.015}}}}
\fontsize{18}{0}
\selectfont\put(311.142,33.0041){\makebox(0,0)[t]{\textcolor[rgb]{0,0,0}{{$V_0$ $[\epsilon]$}}}}
\fontsize{18}{0}
\selectfont\put(42.0014,229.799){\rotatebox{90}{\makebox(0,0)[b]{\textcolor[rgb]{0,0,0}{{$\stackrel[k]{}{\text{mean}}
          \! \bigg\{ \stackrel[\phi]{}{\text{std}}
          \! \big\{ \Delta \alpha_k^\phi \big\} \bigg\}$ $[\ell^{-3}]$}}}}}
\fontsize{18}{0}
\selectfont\put(377.28,103){\makebox(0,0)[t]{\textcolor[rgb]{0,0,0}{{1}}}}
\fontsize{18}{0}
\selectfont\put(391.68,103){\makebox(0,0)[t]{\textcolor[rgb]{0,0,0}{{2}}}}
\fontsize{18}{0}
\selectfont\put(406.08,103){\makebox(0,0)[t]{\textcolor[rgb]{0,0,0}{{3}}}}
\fontsize{18}{0}
\selectfont\put(420.48,103){\makebox(0,0)[t]{\textcolor[rgb]{0,0,0}{{4}}}}
\fontsize{18}{0}
\selectfont\put(434.88,103){\makebox(0,0)[t]{\textcolor[rgb]{0,0,0}{{5}}}}
\fontsize{18}{0}
\selectfont\put(449.28,103){\makebox(0,0)[t]{\textcolor[rgb]{0,0,0}{{6}}}}
\fontsize{18}{0}
\selectfont\put(463.68,103){\makebox(0,0)[t]{\textcolor[rgb]{0,0,0}{{7}}}}
\fontsize{18}{0}
\selectfont\put(478.08,103){\makebox(0,0)[t]{\textcolor[rgb]{0,0,0}{{8}}}}
\fontsize{18}{0}
\selectfont\put(492.48,103){\makebox(0,0)[t]{\textcolor[rgb]{0,0,0}{{9}}}}
\fontsize{18}{0}
\selectfont\put(357.88,108){\makebox(0,0)[r]{\textcolor[rgb]{0,0,0}{{-0.01}}}}
\fontsize{18}{0}
\selectfont\put(357.88,162){\makebox(0,0)[r]{\textcolor[rgb]{0,0,0}{{0}}}}
\fontsize{18}{0}
\selectfont\put(357.88,216){\makebox(0,0)[r]{\textcolor[rgb]{0,0,0}{{0.01}}}}
\fontsize{18}{0}
\selectfont\put(434.88,81){\makebox(0,0)[t]{\textcolor[rgb]{0,0,0}{{$k$}}}}
\fontsize{18}{0}
\selectfont\put(330,162){\rotatebox{90}{\makebox(0,0)[b]{\textcolor[rgb]{0,0,0}{{$\Delta \alpha_k$ $[\ell^{-3}]$}}}}}
\end{picture}
  }
  \caption{Fluctuating potential produces noise on the geometric
    coefficients $\alpha_k$. The results are obtained for the
    $N=10$-particle system with $M=200$ realizations of the
    fluctuating potential for each value of the noise strength $V_0$. 
    The main panel shows the uncertainties (one standard deviation) 
    of $\Delta \alpha_k$, averaged over all $k=1,\dots,9$, versus $V_0$. 
    A fit to the data yields the proportionality factor 
    of $(0.13994 \pm 0.00427)\ell^{-3}\epsilon^{-1}$. 
    The inset shown $\Delta \alpha_k$'s for $V_0 = 0.05 \epsilon$, 
    with error bars corresponding to one standard deviation.}
  \label{fig:meanunc}
\end{figure}

It is instructive to quantify the response of the exchange
coefficients to the fluctuations of the potential. For a fixed value
of the noise strength $V_0$, we denote by $\alpha_k^\phi$ the $k$'th
coefficient for a given realization of the noisy potential with two
particular values for the random phases $\phi_1$ and $\phi_2$.  The
deviation from the noise-free coefficient $\alpha_k$ recorded in
\ref{sec:appendix} is denoted by $\Delta \alpha_k^\phi \equiv \alpha_k
- \alpha_k^\phi$.  Similarly to the random noise introduced directly
into the coefficients \cite{DeChiara2005,petrosyan2010}, the
deviations $\Delta \alpha_k^\phi$ averaged over many ($M=200$)
realizations of the noisy potential, $\Delta \alpha_k \equiv
\stackrel[\phi]{}{\text{mean}} \!  \{ \Delta \alpha_k^\phi \}$, are
approximately zero for all $k$.  In the inset of
Figure~\ref{fig:meanunc} we show the mean deviations $\Delta \alpha_k$
and theirs uncertainties, $\stackrel[\phi]{}{\text{std}} \! \{\Delta
\alpha_k^\phi \}$, for the case of $V_0 = 0.05\epsilon$. Since the
uncertainties are approximately equal for all $k$, we average the
uncertainties over all $k$ and plot the result versus the noise
strength $V_0$ in Figure~\ref{fig:meanunc} main panel. The data follow
a proportionality law with a slope equal to $(0.13994 \pm 0.00427)
\ell^{-3}\epsilon^{-1}$.  This simple relation can indeed be useful as
it provides direct link between the experimental uncertainties and
noise on the coefficients appearing in the effective spin model.

\subsection{Tilting the potential}

\begin{figure}[t]
  \centering
  \resizebox{.75\columnwidth}{!}{
\setlength{\unitlength}{1pt}
\begin{picture}(0,0)
\includegraphics{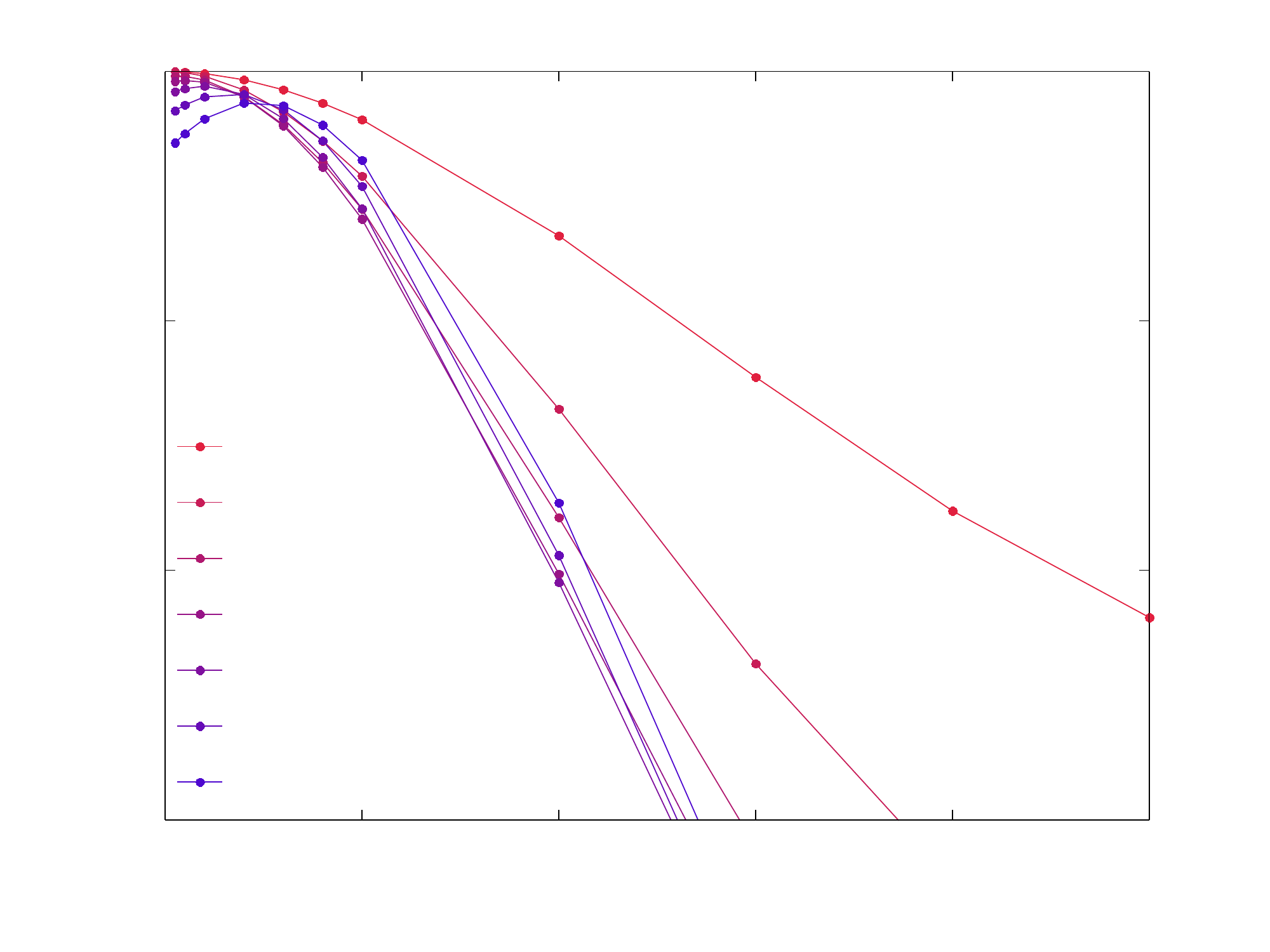}
\end{picture}%
\begin{picture}(576,432)(0,0)
\fontsize{18}{0}
\selectfont\put(74.88,55.0041){\makebox(0,0)[t]{\textcolor[rgb]{0,0,0}{{0}}}}
\fontsize{18}{0}
\selectfont\put(164.16,55.0041){\makebox(0,0)[t]{\textcolor[rgb]{0,0,0}{{0.01}}}}
\fontsize{18}{0}
\selectfont\put(253.44,55.0041){\makebox(0,0)[t]{\textcolor[rgb]{0,0,0}{{0.02}}}}
\fontsize{18}{0}
\selectfont\put(342.72,55.0041){\makebox(0,0)[t]{\textcolor[rgb]{0,0,0}{{0.03}}}}
\fontsize{18}{0}
\selectfont\put(432,55.0041){\makebox(0,0)[t]{\textcolor[rgb]{0,0,0}{{0.04}}}}
\fontsize{18}{0}
\selectfont\put(521.28,55.0041){\makebox(0,0)[t]{\textcolor[rgb]{0,0,0}{{0.05}}}}
\fontsize{18}{0}
\selectfont\put(69.8755,59.9982){\makebox(0,0)[r]{\textcolor[rgb]{0,0,0}{{0.7}}}}
\fontsize{18}{0}
\selectfont\put(69.8755,173.199){\makebox(0,0)[r]{\textcolor[rgb]{0,0,0}{{0.8}}}}
\fontsize{18}{0}
\selectfont\put(69.8755,286.399){\makebox(0,0)[r]{\textcolor[rgb]{0,0,0}{{0.9}}}}
\fontsize{18}{0}
\selectfont\put(69.8755,399.6){\makebox(0,0)[r]{\textcolor[rgb]{0,0,0}{{1}}}}
\fontsize{18}{0}
\selectfont\put(298.08,33.0041){\makebox(0,0)[t]{\textcolor[rgb]{0,0,0}{{$V_0$ $[\epsilon]$}}}}
\fontsize{18}{0}
\selectfont\put(37.8755,229.799){\rotatebox{90}{\makebox(0,0)[b]{\textcolor[rgb]{0,0,0}{{$F(t_\text{out})$}}}}}
\fontsize{18}{0}
\selectfont\put(103.686,229.383){\makebox(0,0)[l]{\textcolor[rgb]{0,0,0}{{$N=4$}}}}
\fontsize{18}{0}
\selectfont\put(103.686,204.011){\makebox(0,0)[l]{\textcolor[rgb]{0,0,0}{{$N=5$}}}}
\fontsize{18}{0}
\selectfont\put(103.686,178.639){\makebox(0,0)[l]{\textcolor[rgb]{0,0,0}{{$N=6$}}}}
\fontsize{18}{0}
\selectfont\put(103.686,153.267){\makebox(0,0)[l]{\textcolor[rgb]{0,0,0}{{$N=7$}}}}
\fontsize{18}{0}
\selectfont\put(103.686,127.895){\makebox(0,0)[l]{\textcolor[rgb]{0,0,0}{{$N=8$}}}}
\fontsize{18}{0}
\selectfont\put(103.686,102.523){\makebox(0,0)[l]{\textcolor[rgb]{0,0,0}{{$N=9$}}}}
\fontsize{18}{0}
\selectfont\put(103.686,77.1514){\makebox(0,0)[l]{\textcolor[rgb]{0,0,0}{{$N=10$}}}}
\end{picture}
  }
  \caption{Retrieval fidelity for $N=10$ particles in a tilted
    potential as a function of tilt amplitude $V_0$ which should be
    compared to the typical energy scale of $0.03\epsilon$.}
  \label{fig:tilt}
\end{figure}
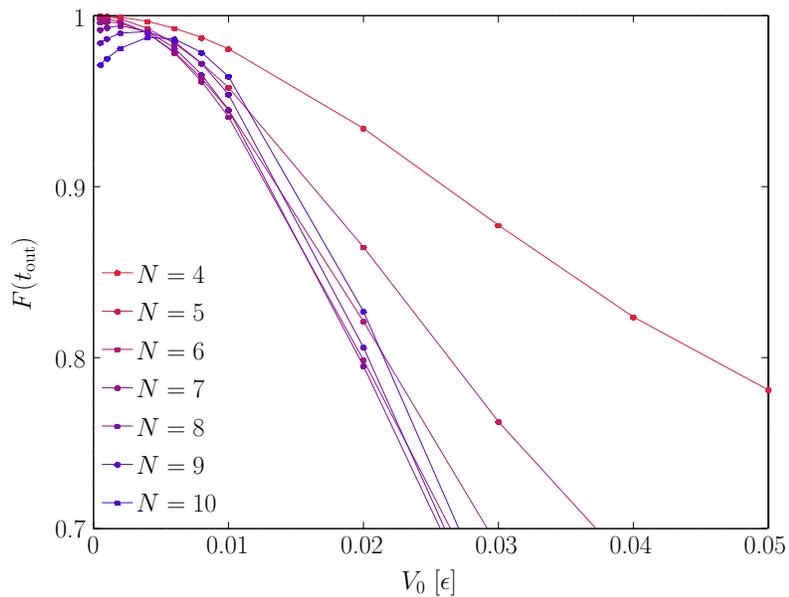

The second kind of experimental uncertainty we consider is tilting the
potential by adding a parity breaking linear term:
\begin{equation}
  \label{eq:tilt}
  \delta V(x) = V_0 \, [x - L/2] / \ell\; ,
\end{equation}
where $V_0$ is a parameter for the tilt amplitude with units of
energy. The value of the slope $V_0/\ell$ should be compared to the
typical energy-over-length scale of the system. An estimate for this
scale for $N=10$ is the quotient between the energy and the spatial
extent of the wavefunction for the 10th single-particle state,
which is $1.21\epsilon / (40 \ell) = 0.03 \epsilon/\ell$. In
Figure~\ref{fig:tilt} we show the state transfer fidelity resulting
from adding the tilt term of Eq. \eref{eq:tilt} to the optimized
potential. We find that the system is tolerant to weak to moderate
tilting noise, $V_0 < 0.005 \epsilon$ (for $N \sim 10$, a sufficiently
small tilt even improves the transfer).  
Larger tilts would strongly suppress state transfer.

\section{Summary}

To conclude, we have employed our efficient computational method
to study the possibility of realizing self-assembled spin chain
systems with strongly-interacting cold atomic gases in one-dimensional
trapping potentials. The parameters of the effective spin chains are
tunable by the confining potential, which can serve as a valuable
experimental tool.

As an important concrete example of quantum dynamics in a many-body
system, we considered quantum state transfer in the effective spin
chains of cold atoms.  We demonstrated how the confining potential
could be tuned to achieve nearly-perfect transfer of a spin excitation
between the two ends of the chain. To emphasize the experimental
feasibility of our proposal, we showed that the cold-atom
implementation of a spin chain is tolerant to moderate noise.
Importantly, we considered noise rigorously by including it directly
in the potential, and not at the level of the effective lattice
Hamiltonian which is only an approximate way of including noise.

In closing we note that throughout this study we have assumed zero 
temperature gas of $N$ strongly-interacting atoms occupying the 
$N$ lowest-energy single-particle eigenstates of the trapping
potential. This approximation is valid for temperatures 
$k_\textrm{B} T \ll \epsilon$, where $k_\textrm{B}$ is the Boltzmann
constant while $\epsilon$ is the typical energy scale of the potential
which characterizes the particle excitation energies. If this
condition is not satisfied, we would have to consider a thermal
distribution of occupation probabilities of different excitation
manifolds of $N$ atoms. For each such manifold, we can construct 
a Slater determinant wavefunction and the corresponding spin chain 
model \cite{volosniev2015}, which in general will have different
local exchange coefficients. Hence, the spin dynamics will experience
decoherence, the more so the higher is the temperature and the 
larger is the number of excited energy manifolds with non-negligible
population. We leave further details for future studies.

\paragraph*{Acknowledgments}
The authors wish to thank A.~G. Volosniev, M. Valiente and
J.~M.~Midtgaard for helpful discussions and comments during the
preparation of the manuscript.  We are grateful to L.~B. Kristensen
and A.~E. Thomsen for development of the computational methods to
determine the geometric coefficients. This work was supported by the
Carlsberg Foundation and by the Danish Council for Independent
Research DFF Natural Sciences and the DFF Sapere Aude program.

\appendix

\section{Geometric coefficients for the state transfer optimized potential}
\label{sec:appendix}
In Section~\ref{sec:perfect} we optimized the $\tau$-dependent
potential of Eq.~\eref{eq:semicirc-potential} such that the geometric
coefficients followed, as close as possible, a semicircle distribution,
leading to perfect state transfer in the spin chain. Here we record 
the optimized values of $\tau$ and the corresponding $N-1$ geometric
coefficients. Since the potential is symmetric, the coefficients have 
the property $\alpha_k = \alpha_{N-k}$, and we only show the values 
of half of them, continuing on the next line as necessary.

\pagebreak 

  \begin{table}[htbp]
    \centering
    \begin{tabular}{l@{\hspace{10pt}}l@{\hspace{10pt}}
        l@{\hspace{9pt}}l@{\hspace{9pt}}l@{\hspace{9pt}}
        l@{\hspace{9pt}}l@{\hspace{9pt}}l@{\hspace{9pt}}}
      $N$ & $\tau$ &$\alpha_1$ &$\alpha_2$ & $\alpha_3$ & $\dots$ &&\\[.4em]
      4  & 2.993540 &0.0475251 &0.0548772 &&&&\\[.2em]
      5  & 3.260198 &0.0510611 &0.0625369 &&&&\\[.2em]
      6  & 3.426142 &0.0579675 &0.0745497 &0.0766875 &&&\\[.2em]
      7  & 3.552248 &0.0658202 &0.0880452 &0.0926727 &&&\\[.2em]
      8  & 3.650098 &0.0743862 &0.1027688 &0.1106852 &0.1116192 &&\\[.2em]
      9  & 3.728815 &0.0834554 &0.1184395 &0.1303163 &0.1326455 &&\\[.2em]
      10 & 3.793287 &0.0929392 &0.1349339 &0.1513421 &0.1557556 &0.1562500 &\\[.2em]
      11 & 3.847014 &0.1027691 &0.1521459 &0.1735853 &0.1807227 &0.1820661 &\\[.2em]
      12 & 3.892360 &0.1128996 &0.1700001 &0.1969183 &0.2073502 &0.2100762 &0.2103721 \\[.2em]
      13 & 3.931069 &0.1232943 &0.1884324 &0.2212349 &0.2354734 &0.2401236 &0.2409746 \\[.2em]
      14 & 3.964425 &0.1339250 &0.2073901 &0.2464477 &0.2649561 &0.2720442 &0.2738566\\
                   &&0.2740494 &&&&&\\[.2em]
      15 & 3.993409 &0.1447681 &0.2268279 &0.2724815 &0.2956824 &0.3056874 &0.3089038\\
                   &&0.3094809 &&&&&\\[.2em]
      16 & 4.018780 &0.1558040 &0.2467065 &0.2992711 &0.3275525 &0.3409190 &0.3459820\\
                   &&0.3472557 &0.3473895 &&&&\\[.2em]
      17 & 4.041134 &0.1670160 &0.2669916 &0.3267593 &0.3604793 &0.3776203 &0.3849582\\
                   &&0.3872880 &0.3877002 &&&&\\[.2em]
      18 & 4.060946 &0.1783897 &0.2876529 &0.3548955 &0.3943860 &0.4156856 &0.4257072\\
                   &&0.4294682 &0.4304027 &0.4305001 &&&\\[.2em]
      19 & 4.078598 &0.1899124 &0.3086636 &0.3836347 &0.4292043 &0.4550208 &0.4681135\\
                   &&0.4736811 &0.4754314 &0.4757379 &&&\\[.2em]
      20 & 4.094403 &0.2015732 &0.3299996 &0.4129363 &0.4648732 &0.4955414 &0.5120715\\
                   &&0.5198130 &0.5226962 &0.5234059 &0.5234795 &&
        \end{tabular}
  \end{table}


\vspace{1cm}

\begin{thebibliography}{99}
\bibitem{lewenstein2007} 
M. Lewenstein, A. Sanpera, V. Ahufinger, B. Damski, A. Sen(De), and U. Sen.
\newblock Ultracold atomic gases in optical lattices: mimicking condensed
  matter physics and beyond.
\newblock {\em Advances in Physics}, 56(2):243--379, 2007.

\bibitem{bloch2008} 
I. Bloch, J. Dalibard, and W. Zwerger.
\newblock Many-body physics with ultracold gases.
\newblock {\em Rev. Mod. Phys.}, 80:885--964, 2008.

\bibitem{esslinger2010} 
T. Esslinger.
\newblock Fermi-Hubbard physics with atoms in an optical lattice.
\newblock {\em Annual Review of Condensed Matter Physics},
1(1):129--152, 2010.

\bibitem{baranov2012} M. A. Baranov, M. Dalmonte, G. Pupillo, and
  P. Zoller.
  \newblock Condensed Matter Theory of Dipolar Quantum Gases.
  \newblock {\em Chem. Rev.}, 112(9):5012--5061, 2012.

\bibitem{zinner2013}
N.~T. Zinner and A.~S. Jensen,
  \newblock Comparing and contrasting nuclei and cold atomic gases.
  \newblock {\em Journal of Physics G: Nuclear and Particle Physics},
  40(5):053101, 2013.

\bibitem{moritz2003} H. Moritz, T. St{\"o}ferle, M. K{\"o}hl, and
  T. Esslinger.
  \newblock Exciting Collective Oscillations in a Trapped 1D Gas.
  \newblock {\em Phys. Rev. Lett.}, 91(25):250402, 2003.

\bibitem{stoferle2004} T. St{\"o}ferle, H. Moritz, C. Schori,
  M. K{\"o}hl, and T. Esslinger.
  \newblock Transition from a Strongly Interacting 1D Superfluid to a Mott Insulator.
  \newblock {\em Phys. Rev. Lett.}, 92(13):130403, 2004.

\bibitem{kinoshita2004} T. Kinoshita, T. Wenger, and D.~S. Weiss.
  \newblock Observation of a One-Dimensional Tonks-Girardeau Gas.
  \newblock {\em Science}, 305(5687):1125--1128, 2004.

\bibitem{paredes2004} B. Paredes {\it et al.}.
  \newblock Tonks–Girardeau gas of ultracold atoms in an optical lattice.
  \newblock {\em Nature}, 429:277--281, 2004.
  
\bibitem{kinoshita2006} T. Kinoshita, T. Wenger, and D.~S. Weiss.
  \newblock A quantum Newton's cradle.
  \newblock {\em Nature}, 440:900--903, 2006.

\bibitem{haller2009} E. Haller {\it et al.}.
  \newblock Realization of an Excited, Strongly Correlated Quantum Gas Phase.
  \newblock {\em Science}, 325(5945):1224--1227, 2009.
  
\bibitem{haller2010} E. Haller {\it et al.}.
  \newblock Pinning quantum  phase transition for a Luttinger liquid
  of strongly interacting bosons.
  \newblock {\em Nature}, 466:597--600, 2010.
  
\bibitem{serwane2011} F. Serwane {\it et al.}.
  \newblock Deterministic Preparation of a Tunable Few-Fermion System.
  \newblock {\em Science}, 332(6027):336--338, 2011.
  
\bibitem{zurn2012} G. Z{\"u}rn {\it et al.}.
  \newblock Fermionization of Two Distinguishable Fermions.
  \newblock {\em Phys. Rev. Lett.}, 108(07):075303, 2012.

\bibitem{wenz2013} A. Wenz {\it et al.}.
  \newblock From Few to Many: Observing the Formation of a Fermi Sea
  One Atom at a Time.
  \newblock {\em Science}, 342(6157):457--460, 2013.
  
\bibitem{pagano2014} G. Pagano {\it et al.}.
  \newblock A one-dimensional liquid of fermions with tunable spin.
  \newblock {\em Nature Phys.}, 10:198--201, 2014.
  
\bibitem{murmann2015} S. Murmann {\it et al.}.
  \newblock Two Fermions in a Double Well: Exploring a Fundamental
  Building Block of the Hubbard Model.
  \newblock {\em Phys. Rev. Lett.}, 114(08):080402, 2015.

\bibitem{chin2010} C. Chin, R. Grimm, P.~S. Julienne, E. and Tiesinga.
  \newblock Feshbach resonances in ultracold gases.
  \newblock {\em Rev. Mod. Phys.}, 82(2):1225, 2010.

\bibitem{olshanii1998} M. Olshanii.
  \newblock Atomic Scattering in the Presence of an External
  Confinement and a Gas of Impenetrable Bosons.
  \newblock {\em Phys. Rev. Lett.}, 81(5):938, 1998.

\bibitem{kuklov2003} A.~B. Kuklov, B.~V. Svistunov.
  \newblock Counterflow Superfluidity of Two-Species Ultracold Atoms
  in a Commensurate Optical Lattice.
  \newblock {\em Phys. Rev. Lett.}, 90(10):100401, 2003.

\bibitem{duan2003} L.-M. Duan, E. Demler, and M.~D. Lukin.
  \newblock Controlling Spin Exchange Interactions of Ultracold Atoms
  in Optical Lattices.
  \newblock {\em Phys. Rev. Lett.}, 91(9):090402, 2003.

\bibitem{fukuhara2013} T. Fukuhara {\it et al.}.
  \newblock Quantum dynamics of a mobile spin impurity.
  \newblock {\em Nature Phys.}, 9:235--241, 2013.

\bibitem{hild2014} S. Hild {\it et al.}.
  \newblock Far-from-Equilibrium Spin Transport in Heisenberg Quantum
  Magnets.
  \newblock {\em Phys. Rev. Lett.}, 113(14):147205, 2014.

\bibitem{volosniev2014} A.~G. Volosniev, D.~V. Fedorov, A.~S. Jensen,
  M. Valiente, and N.~T. Zinner.
  \newblock Strongly interacting confined quantum systems in one
  dimension.
  \newblock {\em Nature Commun.}, 5:5300, 2014.

\bibitem{deuretzbacher2014} F. Deuretzbacher, D. Becker, J. Bjerlin,
  S.~M. Reimann, and L. Santos.
  \newblock Quantum magnetism without lattices in strongly interacting
  one-dimensional spinor gases.
  \newblock {\em  Phys. Rev. A}, 90(1):013611, 2014.

\bibitem{loft2016b} N.~J.~S. Loft, L.~B. Kristensen, A.~E. Thomsen,
  A.~G. Volosniev, and N.~T. Zinner.
  \newblock CONAN -- the cruncher of local exchange coefficients for
  strongly interacting confined systems in one dimension.
  \newblock {\em ArXiv e-prints}, arXiv:1603.02662, 2016.

\bibitem{loft2015} N.~J.~S. Loft, L.~B. Kristensen, A.~E. Thomsen, and
  N.~T. Zinner.
  \newblock Comparing models for the ground state energy of a trapped
  one-dimensional Fermi gas with a single impurity.
  \newblock {\em ArXiv e-prints}, arXiv:1508.05917, 2015.

\bibitem{Deuretzbacher2016}
  F. Deuretzbacher, D. Becker, and L. Santos,
  \newblock Momentum distributions and numerical methods 
  for strongly interacting one-dimensional spinor gases.
  \newblock {\em ArXiv e-prints}, arXiv:1602.06816, 2016

\bibitem{StrRevs2007} S. Bose.
  \newblock Quantum Communication through Spin Chain Dynamics: an
  Introductory Overview.
  \newblock {\em Contemp. Phys.}, 48(1):13--30, 2007;
  
  D. Burgarth.
  \newblock Quantum state transfer and time-dependent disorder in
  quantum chains.
  \newblock {\em Eur. Phys. J. Special Topics}, 151(1):147--155, 2007.

\bibitem{nikolopoulos2014} G. M. Nikolopoulos and I. Jex (eds.).
  \newblock Quantum State Transfer and Network Engineering.
  \newblock Springer, Berlin, 2014.

\bibitem{bose2003} S. Bose.
  \newblock Quantum Communication through an Unmodulated Spin Chain.
  \newblock {\em Phys. Rev. Lett.}, 91(20):207901, 2003.

\bibitem{christandl2004} 
  M. Christandl, N. Datta, A. Ekert, and  A.J. Landahl.
  \newblock Perfect State Transfer in Quantum Spin Networks.
  \newblock {\em Phys. Rev. Lett.}, 92(18):187902, 2004;
  
  M. Christandl, N. Datta, T.C. Dorlas, A. Ekert, A. Kay and A.J. Landahl,
  \newblock Perfect transfer of arbitrary states in quantum spin networks
  \newblock {\em Phys. Rev. A}, 71(3):032312, 2005.

\bibitem{nikolopoulos2004} 
  G.~M. Nikolopoulos, D. Petrosyan, and P. Lambropoulos.
  \newblock Coherent electron wavepacket propagation and entanglement 
  in array of coupled quantum dots.
  \newblock {\em Europhys. Lett.}, 65(3):297, 2004;
 
  G.~M. Nikolopoulos, D. Petrosyan, and P. Lambropoulos.
  \newblock Electron wavepacket propagation in a chain of coupled quantum dots.
  \newblock {\em J. Phys.: Condens. Matter}, 16(28):4991, 2004.

\bibitem{Plenio2004}
M.B. Plenio, J. Hartley, and J. Eisert. 
\newblock Dynamics and manipulation of entanglement in coupled harmonic 
systems with many degrees of freedom. 
\newblock {\em New. J. Phys.} 6:36, 2004.


\bibitem{Bellec2012} M. Bellec, G.M. Nikolopoulos and S. Tzortzakis.
  \newblock Faithful communication Hamiltonian in photonic lattices.
  \newblock {\em Opt. Lett.}, 37(21):4504--4506, 2012.

\bibitem{Perez-Leija2013}
A. Perez-Leija, R. Keil, A. Kay, H. Moya-Cessa, S. Nolte, L.-C. Kwek, 
B. M. Rodriguez-Lara, A. Szameit, and D. N. Christodoulides.
\newblock Coherent quantum transport in photonic lattices.
\newblock {\em Phys. Rev. A}, 87(1):012309, 2013.

\bibitem{volosniev2015} A.~G. Volosniev, D. Petrosyan, M. Valiente,
  D.~V. Fedorov, A.~S. Jensen, and N.~T. Zinner.
  \newblock Engineering the dynamics of effective spin-chain models
  for strongly interacting atomic gases.
  \newblock {\em Phys. Rev. A}, 91(2):023620, 2015.

\bibitem{pu2015} L. Yang, L. Guan, and H. Pu.
  \newblock Strongly interacting quantum gases in one-dimensional traps.
  \newblock {\em Phys. Rev. A}, 91(4):043634, 2015.


\bibitem{levinsen2015} J. Levinsen, P. Massignan, G.~M. Bruun, and
  M.~M. Parish.
  \newblock Strong-coupling ansatz for the one-dimensional Fermi gas
  in a harmonic potential.
  \newblock {\em Science Advances}, 1(6):e1500197, 2015.

\bibitem{hu2015} H. Hu, L. Guan, and S. Chen.
  \newblock  Strongly interacting Bose-Fermi mixtures in one dimension.
  \newblock {\em New J. Phys.}, 18:025009, 2016.

\bibitem{yang2015} L. Yang and X. Cui.
  \newblock Effective spin-chain model for strongly interacting
  one-dimensional atomic gases with an arbitrary spin.
  \newblock {\em Phys. Rev. A}, 93(1):013617, 2016.

\bibitem{yung2006} M.-H. Yung.
  \newblock Quantum speed limit for perfect state transfer in one
  dimension.
  \newblock {\em Phys. Rev. A}, 74(3):030303(R), 2006.

\bibitem{oleksandr2015} O.~V. Marchukov, E.~H. Eriksen,
  J.~M. Midtgaard, A.~A.~S. Kalaee, D.~V. Fedorov, A.~S. Jensen, and
  N.~T. Zinner.
  \newblock Computation of local exchange coefficients in strongly
  interacting one-dimensional few-body systems: local density
  approximation and exact results.
  \newblock {\em Eur. Phys. J. D}, 70:32, 2016.

\bibitem{DeChiara2005} G. De Chiara, D. Rossini, S. Montangero, and
  R. Fazio.
  \newblock From perfect to fractal transmission in spin chains.
  \newblock {\em Phys. Rev. A}, 72(1):012323, 2005.

\bibitem{petrosyan2010} D. Petrosyan, G. M. Nikolopoulos and
  P. Lambropoulos.
  \newblock State transfer in static and dynamic spin chains with
  disorder.
  \newblock {\em Phys. Rev. A}, 81(4):042307, 2010.

\bibitem{Zwick2011} A. Zwick, G.~A. Alvarez, J. Stolze, and O. Osenda.
  \newblock Robustness of spin-coupling distributions for perfect
  quantum state transfer.
  \newblock {\em Phys. Rev. A}, 84(2):022311, 2011.

\bibitem{Nikolopoulos2013} G. M. Nikolopoulos.
  \newblock Statistics of a quantum-state-transfer Hamiltonian in the
  presence of disorder.
  \newblock {\em Phys. Rev. A}, 87(4):042311, 2013.

\bibitem{diener2001} R.~B.~Diener, G.~A. Georgakis, J. Zhong,
  M. Raizen, and Q. Niu.
  \newblock Transition between extended and
  localized states in a one-dimensional incommensurate optical
  lattice.
  \newblock {\em Phys. Rev. A}, 64(3):033416, 2001.


\end{thebibliography}
\end{document}